\documentclass[12pt]{article}
\usepackage{a4wide}
\usepackage{latexsym}
\usepackage{amsmath}
\usepackage{amsfonts}
\usepackage{amscd}
\usepackage{cite}
\usepackage{graphicx}
\usepackage{float}

\usepackage{pslatex}
\usepackage[latin1]{inputenc}
\usepackage[OT2,T1]{fontenc}

\newcommand{\bq}{\begin{eqnarray}}
\newcommand{\eq}{\end{eqnarray}}
\newcommand{\eps}{\varepsilon}

\newcommand{\sla}{\!\!\!/}

\newcommand{\mynosign}{}
\newcommand{\mysign}{-}

\DeclareSymbolFont{cyrletters}{OT2}{wncyr}{m}{n}
\DeclareMathSymbol{\Sha}{\mathalpha}{cyrletters}{"58}

\usepackage{color}

\begin{document}

\thispagestyle{empty}

\begin{flushright}
  MZ-TH/12-20
\end{flushright}

\vspace{1.5cm}

\begin{center}
  {\Large\bf Efficiency improvements for the numerical computation of NLO corrections \\
  }
  \vspace{1cm}
  {\large Sebastian Becker, Christian Reuschle and Stefan Weinzierl\\
\vspace{2mm}
      {\small \em Institut f{\"u}r Physik, Universit{\"a}t Mainz,}\\
      {\small \em D - 55099 Mainz, Germany}\\
  } 
\end{center}

\vspace{2cm}

\begin{abstract}\noindent
  {
In this paper we discuss techniques, which lead to a significant improvement of the efficiency 
of the Monte Carlo integration, when
one-loop QCD amplitudes are calculated numerically with the help of the subtraction method
and contour deformation.
The techniques discussed are:
holomorphic and non-holomorphic division into sub-channels,
optimisation of the integration contour,
improvement of the ultraviolet subtraction terms,
importance sampling and antithetic variates in loop momentum space,
recurrence relations.
   }
\end{abstract}

\vspace*{\fill}

\newpage

\section{Introduction}

In a recent letter \cite{Becker:2011vg} we reported on the next-to-leading order (NLO) results 
in the leading-colour approximation for
five, six and seven jets in electron-positron annihilation.
This was the first time a physical observable depending on a one-loop eight-point function
has been calculated.
The calculation has been performed with a method, which uses numerical Monte Carlo integration for
the computation of the one-loop amplitude.
To render this part finite, 
the subtraction method \cite{Becker:2010ng,Assadsolimani:2010ka,Assadsolimani:2009cz,Nagy:2003qn}
and a contour deformation \cite{Becker:2010ng,Assadsolimani:2010ka,Gong:2008ww,Anastasiou:2007qb,Nagy:2006xy,Soper:2001hu,Soper:1999xk,Soper:1998ye}
have been used.
Also, it has been the first time ever that the numerical method has been applied to a cutting-edge process.
Competing groups use either 
unitarity techniques \cite{Berger:2009zg,Berger:2009ep,Berger:2010vm,Berger:2010zx,Ita:2011wn,Bern:2011ep,Ellis:2009zw,KeithEllis:2009bu,Melia:2010bm,Bevilacqua:2010ve,Bevilacqua:2009zn,Frederix:2010ne}
or Feynman diagram techniques \cite{Bredenstein:2009aj}.
For the unitarity method or the Feynman diagram method a variety of packages are 
available \cite{Hirschi:2011pa,Bevilacqua:2011xh,Cullen:2011kv,Cullen:2011ac,Badger:2010nx,Binoth:2008uq,vanHameren:2010cp,Denner:2010tr,Ossola:2007ax}.

Within all approaches for multi-parton NLO calculations, 
the computation of the virtual corrections is the most challenging part.
The basic principles how the virtual corrections are calculated within the numerical method 
have already been discussed in the 
literature \cite{Becker:2010ng,Assadsolimani:2010ka,Assadsolimani:2009cz,Nagy:2003qn,Gong:2008ww,Anastasiou:2007qb,Nagy:2006xy,Soper:2001hu,Soper:1999xk,Soper:1998ye}.
What is missing in the literature is information on how this can be done efficiently.
Efficiency is a crucial ingredient to apply the method to high multiplicity processes involving seven or eight point functions.

In this paper we provide detailed information how the numerical method for the computation of the virtual corrections can be
implemented in an efficient way.
There are two important parts:
The first essential part is the reduction of the statistical error of the Monte Carlo integration for the virtual part.
We employ several techniques to achieve this goal.
The techniques used are:
holomorphic and non-holomorphic division into sub-channels,
optimisation of the integration contour,
improvement of the ultraviolet subtraction terms
as well as
importance sampling and antithetic variates in loop momentum space.
The numerical method uses contour deformation and subtraction terms to avoid singularities.
In the vicinity of a singularity we use whenever possible contour deformation to by-pass the singularity.
In the case where this is not possible, because the contour is pinched, and if the singularity is not integrable
then there is a subtraction term for it.
This occurs in the soft and collinear regions.
We expect that the regions close to soft and collinear singularities are delicate with respect to the Monte Carlo integration
and will give a significant contribution to the overall Monte Carlo integration error.
This is indeed the case and a solution of this problem belongs in the category of solutions for the
``known unknowns''.
It turned out that we also needed a solution from the category of the ``unknown unknowns''.
In a first attempt we started with an integrand which falls off like $|k|^{-5}$ for $|k|\rightarrow \infty$.
Formally this is sufficient to ensure ultraviolet finiteness.
However, it turned out that the ultraviolet region gave a large contribution to the overall Monte Carlo error.
A significant part of this paper is devoted on the improvement of the ultraviolet behaviour.

The second part is the efficient calculation of the integrand. To achieve this goal, recurrence relations are used.
We show how the integrand of the bare one-loop amplitude and the subtraction
terms can be calculated efficiently.
In essence the calculation of the integrand of the bare one-loop amplitude reduces to a tree-like recurrence relation, once
the loop has been cut open in a single place.

This paper is organised as follows:
In the next section we briefly review the basic principles of the numerical method for
the computation of the virtual corrections.
In section~\ref{sect:monte_carlo} we discuss in detail the techniques which we use to reduce the statistical Monte Carlo error.
Section~\ref{sect:recurrence} is devoted to recurrence relations. 
Finally, section~\ref{sect:conclusions} contains our conclusions.

\section{Numerical calculation of one-loop amplitudes}
\label{sect:general_method}

In this section we briefly outline the method for the numerical computation of NLO corrections.
More details can be found in \cite{Becker:2010ng}.
For concreteness we discuss the case of electron-positron annihilation.
However all methods can equally well be applied to hadron-hadron collisions and deep-inelastic
scattering.

\subsection{Overview of the method}

In electron-positron annihilation 
the contributions at leading and next-to-leading order for an infrared-safe observable $O$ are given as
\bq
\langle O \rangle^{\mathrm{LO}} =  
 \int\limits_n O_n d\sigma^{\mathrm{B}},
 & &
\langle O \rangle^{\mathrm{NLO}} =  
 \int\limits_{n+1} O_{n+1} d\sigma^{\mathrm{R}} + \int\limits_n O_n d\sigma^{\mathrm{V}}.
\eq
Here a rather condensed notation is used. $d\sigma^{\mathrm{B}}$ denotes the Born
contribution,
whose matrix elements are given by the square of the Born amplitudes with $(n+2)$ particles.
Similar, $d\sigma^{\mathrm{R}}$ denotes the real emission contribution,
whose matrix elements are given by the square of the Born amplitudes with $(n+3)$ particles.
$d\sigma^{\mathrm{V}}$ gives the virtual contribution, whose matrix elements are given by the interference term
of the one-loop amplitude with $(n+2)$ particles, with the corresponding
Born amplitude.
Taken separately, the individual contributions at next-to-leading order 
are divergent and only their sum is finite.
In order to render the real emission contribution finite, such that the phase space integration
can be performed by Monte Carlo methods, one adds and subtracts a suitably chosen piece
\cite{Catani:1997vz,Dittmaier:1999mb,Phaf:2001gc,Catani:2002hc}:
\bq
\langle O \rangle^{\mathrm{NLO}} & = & 
 \int\limits_{n+1} \left( O_{n+1} d\sigma^{\mathrm{R}} - O_n d\sigma^{\mathrm{A}} \right)
 + \int\limits_n \left( O_n d\sigma^{\mathrm{V}} + O_n \int\limits_1 d\sigma^{\mathrm{A}} \right).
\eq
The term $(O_{n+1}  d\sigma^{\mathrm{R}} - O_n d\sigma^{\mathrm{A}})$ in the first bracket is by construction integrable over the
$(n+1)$-particle phase space and can be evaluated numerically.
The subtraction term can be integrated analytically over the unresolved one-particle phase space.
In a compact notation the result of this integration is often written as
\bq
 \int\limits_1 d\sigma^{\mathrm{A}}  
 & = & {\bf I} \otimes d\sigma^{\mathrm{B}}.
\eq
The notation $\otimes$ indicates that colour correlations due to the colour charge operators
still remain.
The virtual contribution
$d\sigma^{\mathrm{V}}$ is given by
\bq
 d\sigma^{\mathrm{V}} & = & 2 \;\mbox{Re}\; \left(\left.{\cal A}^{(0)}\right.^\ast {\cal A}^{(1)} \right) d\phi_n.
\eq
${\cal A}^{(1)}$ denotes the renormalised one-loop amplitude. It is related to the bare amplitude by
\bq
\label{eq_one_loop}
 {\cal A}^{(1)} & = & {\cal A}^{(1)}_{\mathrm{bare}} + {\cal A}^{(1)}_{\mathrm{CT}}.
\eq
${\cal A}^{(1)}_{\mathrm{CT}}$ denotes the ultraviolet counterterm from renormalisation.
The bare one-loop amplitude involves the loop integration
\bq
\label{integrand_one_loop}
{\cal A}^{(1)}_{\mathrm{bare}} & = & \int \frac{d^Dk}{(2\pi)^D} {\cal G}^{(1)}_{\mathrm{bare}},
\eq
where ${\cal G}^{(1)}_{\mathrm{bare}}$ denotes the integrand of the bare one-loop amplitude.
Within our approach we extend the subtraction method to the integration over the virtual particles circulating in the loop.
To this aim we rewrite eq.~(\ref{eq_one_loop}) as
\bq
\label{basic_subtraction_loop}
 {\cal A}_{\mathrm{bare}}^{(1)} + {\cal A}_{\mathrm{CT}}^{(1)} 
 & = & 
 \left( {\cal A}_{\mathrm{bare}}^{(1)} - {\cal A}_{\mathrm{soft}}^{(1)} - {\cal A}_{\mathrm{coll}}^{(1)} - {\cal A}_{\mathrm{UV}}^{(1)} \right)
 + \left( {\cal A}_{\mathrm{CT}}^{(1)}  
 + {\cal A}_{\mathrm{soft}}^{(1)} + {\cal A}_{\mathrm{coll}}^{(1)} + {\cal A}_{\mathrm{UV}}^{(1)} \right).
\eq
The subtraction terms ${\cal A}_{\mathrm{soft}}^{(1)}$, ${\cal A}_{\mathrm{coll}}^{(1)}$ and ${\cal A}_{\mathrm{UV}}^{(1)}$
are chosen such that they match locally the singular behaviour of the integrand of ${\cal A}_{\mathrm{bare}}^{(1)}$ in $D$ dimensions.
The first bracket in eq.~(\ref{basic_subtraction_loop})
can therefore be integrated numerically in four dimensions.
The term ${\cal A}_{\mathrm{soft}}^{(1)}$ approximates the soft singularities, 
${\cal A}_{\mathrm{coll}}^{(1)}$ approximates
the collinear singularities and 
the term ${\cal A}_{\mathrm{UV}}^{(1)}$ approximates the ultraviolet singularities.
These subtraction terms have a local form similar to eq.~(\ref{integrand_one_loop}):
\bq
{\cal A}^{(1)}_{\mathrm{soft}} = \int \frac{d^Dk}{(2\pi)^D} {\cal G}^{(1)}_{\mathrm{soft}},
\;\;\;\;\;\;
{\cal A}^{(1)}_{\mathrm{coll}} = \int \frac{d^Dk}{(2\pi)^D} {\cal G}^{(1)}_{\mathrm{coll}},
\;\;\;\;\;\;
{\cal A}^{(1)}_{\mathrm{UV}} = \int \frac{d^Dk}{(2\pi)^D} {\cal G}^{(1)}_{\mathrm{UV}}.
\eq
The contribution from the terms in the first bracket of eq.~(\ref{basic_subtraction_loop})
can be written as
\bq
\label{eq_monte_carlo_integration}
\lefteqn{
 \int
 2 \;\mbox{Re}\; \left[\left.{\cal A}^{(0)}\right.^\ast 
   \left( {\cal A}_{\mathrm{bare}}^{(1)} - {\cal A}_{\mathrm{soft}}^{(1)} - {\cal A}_{\mathrm{coll}}^{(1)} - {\cal A}_{\mathrm{UV}}^{(1)} \right) 
                \right]{\cal O}_n d\phi_n
 = } & &
 \nonumber \\
 & & 
 \int d\phi_n \int \frac{d^4k}{(2\pi)^4} 
 2 \;\mbox{Re}\; 
 \left[\left.{\cal A}^{(0)}\right.^\ast 
   \left( {\cal G}_{\mathrm{bare}}^{(1)} - {\cal G}_{\mathrm{soft}}^{(1)} 
        - {\cal G}_{\mathrm{coll}}^{(1)} - {\cal G}_{\mathrm{UV}}^{(1)} \right) 
                \right]
 {\cal O}_n
 + {\cal O}\left(\eps\right).
\eq
The integral on the right-hand side is finite.
Note that the integration over the loop momentum $k$ can be done together with the phase space integration in a single Monte Carlo
integration.
The building blocks of the subtraction terms are process-independent. When adding them back, we integrate analytically
over the loop momentum $k$. The result can be written as
\bq
 2 \;\mbox{Re}\; \left[\left.{\cal A}^{(0)}\right.^\ast 
   \left( {\cal A}_{\mathrm{CT}}^{(1)} + {\cal A}_{\mathrm{soft}}^{(1)} + {\cal A}_{\mathrm{coll}}^{(1)} + {\cal A}_{\mathrm{UV}}^{(1)}\right) 
                \right] d\phi_n
 & = & 
 {\bf L} \otimes d\sigma^{\mathrm{B}}.
\eq
The insertion operator ${\bf L}$ contains the explicit poles in the dimensional regularisation parameter related
to the infrared singularities of the one-loop amplitude.
These poles cancel when combined with the insertion operator ${\bf I}$:
\bq
 \left( {\bf I} + {\bf L} \right) \otimes d\sigma^{\mathrm{B}} & = &
 \mbox{finite}.
\eq
For massless particles we have
\bq
 {\bf I} + {\bf L}
 & = &
 \frac{\alpha_s}{2\pi}
 \; \mbox{Re}
 \left[
  \sum\limits_i \sum\limits_{j\neq i} {\bf T}_i {\bf T}_j 
      \left( 
              \frac{\gamma_i}{{\bf T}_i^2} \ln \frac{\left|2p_ip_j\right|}{\mu_{\mathrm{UV}}^2}
            - \frac{\pi^2}{2} \theta(2p_ip_j)
      \right)
 \right. \nonumber \\
 & & \left.
 + \sum\limits_i \left( \gamma_i + K_i - \frac{\pi^2}{3} {\bf T}_i^2 \right)
 -  \frac{(n-2)}{2} \beta_0 \ln \frac{\mu_{\mathrm{UV}}^2}{\mu^2}
 \right]
 + {\cal O}(\eps).
\eq
In this formula we denote by $\mu$ the renormalisation scale and by $\mu_{\mathrm{UV}}$ the scale used in the
ultraviolet subtraction terms.
Of course it is possible to set $\mu=\mu_{\mathrm{UV}}$, but it will be advantageous to keep this two scales different.
In fact, we will choose $\mu_{\mathrm{UV}}^2$ purely imaginary.
The colour charge operator for particle $i$ is denoted by ${\bf T}_i$. We further have
\bq
 {\bf T}_q^2 = {\bf T}_{\bar{q}}^2 = C_F, 
 & &
 {\bf T}_g^2 = C_A, 
 \nonumber \\
 \gamma_q = \gamma_{\bar q} = \frac{3}{2} C_F, 
 & &
 \gamma_g = \frac{1}{2} \beta_0,
 \nonumber \\
 K_q = K_{\bar q} = \left( \frac{7}{2} - \frac{\pi^2}{6} \right) C_F,
 & &
 K_g = \left( \frac{67}{18} - \frac{\pi^2}{6} \right) C_A - \frac{10}{9} T_R N_f.
\eq
$\beta_0$ is the first coefficient of the QCD $\beta$-function and given by
\bq
\beta_0 = \frac{11}{3} C_A - \frac{4}{3} T_R N_f.
\eq
The colour factors are as usual
\bq
C_A = N_c, \;\;\; C_F = \frac{N_c^2-1}{2N_c}, \;\;\; T_R = \frac{1}{2}.
\eq
We therefore have to evaluate three contributions to get the next-to-leading order correction:
The real emission contribution
\bq
\langle O \rangle^{\mathrm{NLO}}_{\mathrm{real}} & = & 
 \int\limits_{n+1} \left( O_{n+1} d\sigma^{\mathrm{R}} - O_n d\sigma^{\mathrm{A}} \right),
\eq
the insertion term
\bq
\langle O \rangle^{\mathrm{NLO}}_{\mathrm{insertion}} & = & 
 \int\limits_{n} O_{n} \left( {\bf I} + {\bf L} \right) \otimes d\sigma^{\mathrm{B}},
\eq
and the virtual contribution
\bq
\label{eq_virtual}
\langle O \rangle^{\mathrm{NLO}}_{\mathrm{virtual}} & = & 
 2 \int d\phi_n \;\mbox{Re}\; \int \frac{d^4k}{(2\pi)^4}
 \left[\left.{\cal A}^{(0)}\right.^\ast 
 \left( {\cal G}_{\mathrm{bare}}^{(1)} - {\cal G}_{\mathrm{soft}}^{(1)} 
        - {\cal G}_{\mathrm{coll}}^{(1)} - {\cal G}_{\mathrm{UV}}^{(1)} \right) \right] O_{n} .
\eq
All three contributions are by construction finite.
In this paper we focus on the virtual contribution
$\langle O \rangle^{\mathrm{NLO}}_{\mathrm{virtual}}$.

\subsection{Colour decomposition and kinematics}

It is convenient to decompose a full one-loop QCD amplitude into primitive amplitudes:
\bq
 {\cal A}^{(1)} & = & \sum\limits_{j} C_j A^{(1)}_j.
\eq
The colour structures are denoted by $C_j$ while
the primitive amplitudes are denoted by $A^{(1)}_j$.
In the colour-flow basis \cite{'tHooft:1973jz,Maltoni:2002mq,Weinzierl:2005dd} the colour structures are linear combinations of monomials in Kronecker $\delta_{ij}$'s.
A primitive amplitudes is defined as a colour-stripped 
gauge-invariant set of Feynman diagrams with a fixed cyclic ordering of
the external partons and a definite routing 
of the external fermion lines through the diagram \cite{Bern:1994fz}.

It is simpler to work with primitive one-loop amplitudes instead of a full one-loop amplitude.
Our method exploits the fact that primitive one-loop amplitudes have a fixed cyclic ordering of the
external legs and that they are gauge-invariant.
The first point ensures that there are at maximum $n$ different loop propagators in the problem, where $n$ is the 
number of external legs, while the second property of gauge invariance is crucial for the proof of the
method.
We therefore consider in the following just a single primitive one-loop amplitude,
which we denote by $A^{(1)}$, while keeping in mind that the full one-loop amplitude is just
the sum of several primitive amplitudes multiplied by colour structures.
The full one-loop amplitude is obtained by summing over all primitive amplitudes.

We introduce some notation related to the primitive amplitude $A^{(1)}$.
Since the cyclic ordering of the external particles is fixed, 
there are only $n$ different propagators occurring
in the loop integral.
We label the external momenta clockwise by $p_1$, $p_2$, ..., $p_n$ 
and define $q_i=p_1+p_2+...+p_i$.
The loop momenta are
\bq
 k_j & = & k - q_j, 
 \;\;\;
 q_j = \sum\limits_{l=1}^j p_l.
\eq
For convenience we set
\bq
 k_0 = k_n & \mbox{and} & q_0 = q_n.
\eq
Due to momentum conservation we actually have
\bq
 q_0 = q_n = 0.
\eq
Nevertheless we will use $q_0$ (or $q_n$), this makes the formulae more symmetric with respect to the indices.
In electron-positron annihilation we can take $p_1$, $p_2$, ..., $p_{n-2}$ to be the final state momenta and
$p_{n-1}$ and $p_n$ to be the (negative) of the initial state momenta.
The two leptons couple only through a photon or through a $Z$-boson to the quark line.
This implies that in primitive amplitudes related to electron-positron annihilation only
$(n-1)$ loop propagators are present.
Phrased differently, 
in electron-positron annihilation the propagator corresponding to the loop momentum $k_{n-1}$ is absent.
We can write the bare primitive one-loop amplitude in Feynman gauge as
\bq
\label{starting_point}
 A^{(1)}_{\mathrm{bare}} = \int \frac{d^Dk}{(2\pi)^D} 
 G^{(1)}_{\mathrm{bare}},
 & &
 G^{(1)}_{\mathrm{bare}} = 
 P_{\mathrm{bare}}(k) \prod\limits_{i=0}^{n-2} \frac{1}{k_i^2 - m_i^2 + i \delta}.
\eq
$G^{(1)}_{\mathrm{bare}}$ is the integrand of the bare one-loop amplitude.
$P_{\mathrm{bare}}(k)$ is a polynomial in the loop momentum $k$.
The $+i\delta$-prescription instructs us to deform -- if possible -- the integration contour into the complex plane
to avoid the poles at $k_i^2=m_i^2$.
In the following we do not write the $+i\delta$-term explicitly.

\subsection{The subtraction terms}

At the level of primitive amplitudes we denote the subtraction terms for the integrand,
corresponding to the ones appearing in eq.~(\ref{eq_virtual}) by
\bq
G^{(1)}_{\mathrm{soft}}, 
 \;\;\;
G^{(1)}_{\mathrm{coll}}, 
 \;\;\;
G^{(1)}_{\mathrm{UV}}. 
\eq
For massless QCD the soft and collinear subtraction terms are given by
\bq
\label{unintegrated_subtraction_terms}
G_{\mathrm{soft}}^{(1)} & = & 
4i\sum\limits_{j\in I_g} \frac{p_{j}.p_{j+1}}{k_{j-1}^2k_{j}^2k_{j+1}^2} A_{j}^{(0)} \;, \nonumber \\
G_{\mathrm{coll}}^{(1)}
& = & -2i\sum\limits_{j\in I_g} \bigg[ \frac{S_{j}g_{\mathrm{UV}}(k_{j-1}^2,k_{j}^2)}{k_{j-1}^2k_{j}^2}
                                    + \frac{S_{j+1}g_{\mathrm{UV}}(k_{j}^2,k_{j+1}^2)}{k_{j}^2k_{j+1}^2} \bigg] A_{j}^{(0)} \;,
\eq
where the sum over $j \in I_g$ is over all gluon propagators $j$ inside the loop. 
If we take the subset of diagrams which have the gluon loop propagator $j$ and if we remove from each diagram of this subset 
the loop propagator $j$ we obtain a set of tree diagrams.
After removing multiple copies of identical diagrams this set 
forms a Born partial amplitude which we denote by $A^{(0)}_{j}$.
Furthermore, $S_{j}=1$ if the external line $j$ corresponds to a quark 
and $S_{j}=1/2$ if it corresponds to a gluon. 
The function $g_{\mathrm{UV}}$ ensures that the integration over the loop momentum is ultraviolet finite.
The function $g_{\mathrm{UV}}$ must have the properties
\bq
\label{requirement_g_UV}
 \lim\limits_{k_{j-1} || k_j} g_{\mathrm{UV}}\left(k_{j-1}^2,k_j^2\right) = 1,
 & & 
 \lim\limits_{k \rightarrow \infty} g_{\mathrm{UV}}\left(k_{j-1}^2,k_j^2\right) 
 = {\cal O}\left( \frac{1}{k} \right).
\eq
A possible choice is \cite{Becker:2010ng}
\bq
\label{def1_g_UV}
 g_{\mathrm{UV}}\left(k_{j-1}^2,k_j^2\right)
 & = &
 1 - \frac{k_{j-1}^2 k_j^2}{\left[ (k-Q)^2-\mu_{\mathrm{UV}}^2 \right]^2}.
\eq
$Q$ is an arbitrary four-vector independent of the loop momentum $k$ and 
$\mu_{\mathrm{UV}}$ is an arbitrary scale. 
Since these two quantities are arbitrary, there are no restrictions on them, they even may have
complex values.
We will later choose $\mu_{\mathrm{UV}}^2$ purely imaginary with 
$\mbox{Im}\;\mu_{\mathrm{UV}}^2 < 0$.
This will ensure that the denominator of eq.~(\ref{def1_g_UV}) does not introduce additional
singularities for the contour integration.
There are many possible choices for the function $g_{\mathrm{UV}}$ compatible with eq.~(\ref{requirement_g_UV}), 
and we will choose an improved form
which minimises the Monte Carlo integration error.
This will be discussed in detail in section~\ref{sect:monte_carlo}.
The soft and collinear subtraction terms for QCD amplitudes with massive partons are also 
known \cite{Becker:2010ng,Assadsolimani:2010ka,Assadsolimani:2009cz}.
The soft and collinear subtraction terms are formulated at the amplitude level and
are proportional to the corresponding Born amplitudes. 
Upon integration they yield simple analytic results:
\bq
S_{\eps}^{-1}\mu^{2\eps}\int \frac{d^Dk}{(2\pi)^D} \, G_{\mathrm{soft}}^{(1)} & = &
-\frac{1}{(4\pi)^2}\frac{e^{\eps\gamma_E}}{\Gamma(1-\eps)}\sum\limits_{j\in I_g} \frac{2}{\eps^2} \Big( \frac{-2p_{j}p_{j+1}}{\mu^2}\Big)^{-\eps} A_j^{(0)} \; + {\cal O}(\eps),
 \nonumber \\
S_{\eps}^{-1}\mu^{2\eps}\int \frac{d^Dk}{(2\pi)^D} \, G_{\mathrm{coll}}^{(1)} & = &
-\frac{1}{(4\pi)^2}\frac{e^{\eps\gamma_E}}{\Gamma(1-\eps)}\sum\limits_{j\in I_g} (S_j+S_{j+1}) \frac{2}{\eps} \Big( \frac{\mu_{\mathrm{UV}}^2}{\mu^2}\Big)^{-\eps} A_j^{(0)} \; + {\cal O}(\eps),
\eq
with $S_{\eps}\equiv(4\pi)^\eps\exp(-\eps\gamma_E)$ the typical volume factor in dimensional regularisation, 
where $\gamma_E$ denotes the Euler-Mascheroni constant, $\mu$ denotes the renormalisation scale in dimensional regularisation 
and $\eps$ is defined through $D = 4-2\eps$.

The ultraviolet subtraction terms correspond to propagator and vertex corrections.
The subtraction terms are obtained by expanding the relevant loop propagators
around a new ultraviolet propagator $(\bar{k}^2-\mu_{\mathrm{UV}}^2)^{-1}$, where $\bar{k} = k - Q$:
For a single propagator we have
\bq
\label{prop_uv_expansion}
 \frac{1}{\left(k-p\right)^2-m^2}
  & = & 
 \frac{1}{\bar{k}^2-\mu_{\mathrm{UV}}^2}
       + \frac{2\bar{k}\cdot\left(p-Q\right)}{\left(\bar{k}^2-\mu_{\mathrm{UV}}^2\right)^2}
 - \frac{\left(p-Q\right)^2-m^2+\mu_{\mathrm{UV}}^2}{\left(\bar{k}^2-\mu_{\mathrm{UV}}^2\right)^2}
 + \frac{\left[ 2\bar{k}\cdot\left(p-Q\right)\right]^2}{\left(\bar{k}^2-\mu_{\mathrm{UV}}^2\right)^3}
 \nonumber \\
 & &
 + {\cal O}\left(\frac{1}{|\bar{k}|^5}\right).
\eq
We can always add finite terms to the subtraction terms.
For the ultraviolet subtraction terms we choose the finite terms such that
the finite parts of the 
integrated ultraviolet subtraction terms
are independent of $Q$ and proportional to the pole part, 
with the same constant of proportionality.
The integrated ultraviolet subtraction terms have the form
\bq
 c \left( \frac{1}{\eps} - \ln \frac{\mu_{\mathrm{UV}}^2}{\mu^2} \right) + {\cal O}(\eps),
\eq
where $c$ depends on the type of the subtraction term.
This ensures that the sum of all integrated UV subtraction terms is again proportional to a tree-level amplitude.

Putting everything together we have
\bq
\label{integrand_form}
G^{(1)}_{\mathrm{bare}} - G^{(1)}_{\mathrm{soft}} - G^{(1)}_{\mathrm{coll}} - G^{(1)}_{\mathrm{UV}} 
 & = &
 \frac{R(k)}{\prod\limits_{j=0}^{n-2} \left( k_j^2 - m_j^2 \right)},
 \nonumber \\
 R(k) & = &
  P(k)
  - \frac{P_{\mathrm{UV}}(k)}{\left( \bar{k}^2 - \mu_{\mathrm{UV}}^2 \right)^{n_{\mathrm{UV}}}} 
    \prod\limits_{j=0}^{n-2} \left( k_j^2 - m_j^2 \right).
\eq
$P(k)$ and $P_{\mathrm{UV}}(k)$ are polynomials in the loop momentum $k$.
$G^{(1)}_{\mathrm{bare}}$ and $G^{(1)}_{\mathrm{soft}}$ defined as in eq.~(\ref{unintegrated_subtraction_terms}) 
contribute only to the first term of the
r.h.s of eq.~(\ref{integrand_form}), while $G^{(1)}_{\mathrm{UV}}$ contributes only to the second 
term on the r.h.s of eq.~(\ref{integrand_form}).
On the other hand, $G^{(1)}_{\mathrm{coll}}$ contributes to both terms on the r.h.s of eq.~(\ref{integrand_form}).

\subsection{The contour deformation}

Having a complete list of ultraviolet and infrared subtraction terms at hand, 
we can ensure that the integration
over the loop momentum gives a finite result and can therefore be performed in four dimensions.
However, this does not yet imply that we can safely integrate each of the four components 
of the loop momentum $k^\mu$
from minus infinity to plus infinity along the real axis.
There is still the possibility that some of the loop propagators go on-shell 
for real values of the loop momentum.
If the contour is not pinched this is harmless, as we may escape into the complex plane 
in a direction indicated by
Feynman's $+i\delta$-prescription.
However, it implies that the integration should be done over a region of real dimension $4$ 
in the complex space ${\mathbb C}^4$.
If the contour is pinched then the singularity is integrable when the integration is done over the loop momentum space and the phase space.
This is the case because either the singularity in the bare one-loop amplitude is integrable by itself, or -- if not -- there is a subtraction term
for it.
Let us look again at eq.~(\ref{integrand_form}).
Choosing $\mu_{\mathrm{UV}}^2$ sufficiently large on the negative imaginary axis ensures that
the integration contour stays always away from the poles defined by $\bar{k}^2 - \mu_{\mathrm{UV}}^2=0$.
Therefore we have to consider for the contour deformation only the poles defined by
$k_j^2 - m_j^2=0$.
The integral which we will have to consider is given by
\bq
\label{complex_integral}
 \int\frac{d^{4}k}{(2\pi)^{4}}
 \left(
       G^{(1)}_{\mathrm{bare}} - G^{(1)}_{\mathrm{soft}} - G^{(1)}_{\mathrm{coll}} - G^{(1)}_{\mathrm{UV}} 
 \right) 
 & = & 
 \int\frac{d^{4}k}{(2\pi)^{4}}
 \frac{R(k)}{\prod\limits_{j=0}^{n-2} \left( k_j^2 - m_j^2 \right)},
\eq
where $R(k)$ is a rational function of the loop momentum $k^{\mu}$, which has only poles at $\bar{k}^2-\mu_{\mathrm{UV}}^2=0$ 
and the integration is over a complex contour in order to avoid 
whenever possible the poles of the propagators shown explicitly in eq.~(\ref{complex_integral}). 

Let us first look for conditions how the integration contour has to be chosen.
We set 
\bq
 k & = & 
 \tilde{k} + i \kappa(\tilde{k}),
\eq 
where $\tilde{k}^{\mu}$ is real. After this deformation our integral equals
\bq
 & &
 \int\frac{d^{4}\tilde{k}}{(2\pi)^{4}}
 \left|\frac{\partial k^{\mu}}{\partial \tilde{k}^{\nu}}\right|
 \frac{R(k(\tilde{k}))}{\prod\limits_{j=0}^{n-2} \left(\tilde{k}_{j}^{2}-m_{j}^{2}-\kappa^{2}+2i \tilde{k}_{j}\cdot\kappa \right)}.
\eq
To match Feynman's $+i\delta$-prescription
we have to construct the deformation vector $\kappa$ such that
\bq
\tilde{k}_{j}^{2}-m_{j}^{2} & = & 0 \quad \rightarrow \quad \tilde{k}_{j}\cdot \kappa \geq 0,
\eq
and the equal sign applies only if the contour is pinched.
Integration contours, which fulfil these requirements are called allowed contours.
Among all allowed contours we would like to pick one, 
\begin{itemize}
\item which we can construct algorithmically in a process-independent way and
\item which leads to a small Monte Carlo integration error.
\end{itemize}
The first requirement is clearly needed for a process-independent method, the second requirement is essential
for the efficiency of the method.
We have implemented and tested two algorithms for the contour deformation.
The first one uses an auxiliary Feynman parametrisation \cite{Nagy:2006xy,Becker:2010ng}, where also the Feynman parameters
are deformed into the complex Feynman parameter space.
The Monte Carlo integration is then over the phase space of the final state particles, the loop momentum space and the Feynman parameter space.
The second algorithm follows ref.~\cite{Gong:2008ww} and works directly in loop momentum space. 
In this case the Monte Carlo integration is only over the final state phase space and the loop momentum space.

The advantages and disadvantages of the two algorithms are as follows:
Within the algorithm based on Feynman parametrisation the definition of the integration contours is straightforward,
even in the case of massive propagators.
After Feynman parametrisation the singularities in loop momentum space are for fixed external momenta and fixed Feynman parameters
on a single cone. The deformation for the Feynman parameters follows from a kinematical matrix, which depends only on the external invariants.
However, all terms including integrable singularities are raised to the power $n$, being the number of external particles.
This is an artefact of Feynman integration. For non-degenerate external kinematics we can have in four space-time dimensions maximally four
propagators which go simultaneously on-shell.
The additional integration over the Feynman parameters will effectively lower the degree of the singularities down to the correct one.
However, this integration is done numerically and is therefore a source of large Monte Carlo errors.

Within the algorithm, which works directly in loop momentum space, the deformation for the integration contour is more involved and currently only known in the massless case,
but it has the advantage that this algorithm does not artificially enhance the degree of a singularity.
Within this algorithm the singularities lie on the cones $(k-q_i)^2=0$ with origins given by $q_0$, $q_1$, ..., $q_{n-1}$.
\begin{figure}
\begin{center}
\includegraphics[bb=90 600 515 735,width=0.9\textwidth]{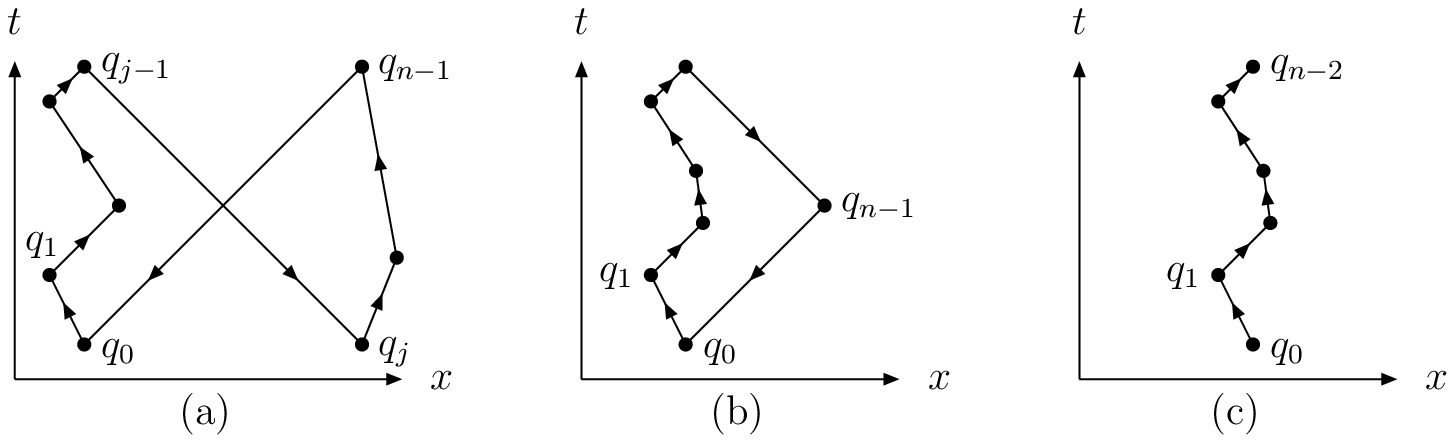}
\caption{\label{fig_zig_zag}
A sketch of the loop momentum space: Diagram $(a)$ corresponds to a generic
primitive amplitude, diagram $(b)$ corresponds to a primitive amplitude, where
the two incoming particles are adjacent, diagram $(c)$ corresponds to the situation
in electron-positron annihilation, where the poles due to $q_{n-1}$ are absent.
}
\end{center}
\end{figure}
Since
\bq 
 p_i & = & q_i - q_{i-1},
\eq
the external momenta $p_i$ connect the origins of the cones and we arrive 
for a generic primitive one-loop amplitude at the
graphical representation shown in fig.(\ref{fig_zig_zag}) in diagram $(a)$.
This is the graphical representation of a Wilson loop.
For two initial state particles we always have two strands 
in the positive $t$-direction.
This is shown in diagram $(a)$ where we have one strand from $q_0$ to $q_{j-1}$
and another strand from $q_{j}$ to $q_{n-1}$.
In the case, where the two initial state momenta are adjacent, the diagram
degenerates to the one shown in diagram $(b)$ in fig.~(\ref{fig_zig_zag}).
If in addition, particle $(n-1)$ and particle $n$ always couple
through an intermediate particle (e.g. photon or $Z$-boson) to the loop, then
the propagator corresponding to $q_{n-1}$ is absent in the loop and we obtain the situation shown
in diagram $(c)$ in fig.~(\ref{fig_zig_zag}) consisting of a single strand.

In loop momentum space the points $q_0$, $q_1$, ..., $q_{n-1}$ are contained within a finite region, which we call the
``interior region''.
In diagram $(c)$ of fig.~\ref{fig_zig_zag} we may define the interior region by the 
intersection of the interior of the backward light cone from $q_{n-2}$ with the interior of the forward light cone from $q_0$.
The complement to the interior region we call the ``exterior region''.

Both algorithms yield identical results, with the direct loop momentum deformation method being more efficient for processes with a large number
of external particles.
The algorithm based on Feynman parameters has already been described together with optimisation techniques in ref.~\cite{Becker:2010ng}.
In this paper we focus on the direct loop momentum deformation method and discuss in the next section optimisation techniques
for this method.


\section{Optimisation techniques}
\label{sect:monte_carlo}

In this section we discuss the employed optimisation techniques for the Monte Carlo integration over the loop momentum space 
and the phase space of the final state particles.
The contour of integration is deformed directly in loop momentum space, without the introduction of Feynman parameters.
The optimisation techniques are a combination of 
standard Monte Carlo optimisation techniques \cite{Weinzierl:2000wd}
(which can be applied to any Monte Carlo integration and do not require any particular information on the integrand),
and improvements obtained by taking into account the physical nature of the problem.
We point out that in the latter case we only use information which follows from the general principles of quantum field theory, 
for example that the singularities of the integrand are given by the poles of the propagators.
We do not use any process-dependent information.
Therefore, all optimisation techniques are process-independent.

\subsection{Holomorphic and non-holomorphic division into sub-channels}

We start with a general observation. Consider a complex contour integral
\bq
 I & = & \int\limits_{\cal C} dz \; f(z),
\eq
where $f(z)$ is a meromorphic function.
Suppose $f(z)=f_1(z)+f_2(z)$, where $f_1(z)$ and $f_2(z)$ are again meromorphic functions of $z$.
Then
\bq
 I & = &
 \int\limits_{{\cal C}_1} dz \; f_1(z)
 +
 \int\limits_{{\cal C}_2} dz \; f_2(z).
\eq
The contours ${\cal C}_1$ and ${\cal C}_2$ may differ from the original contour ${\cal C}$, as long as no poles are crossed in going from
${\cal C}$ to ${\cal C}_1$ or ${\cal C}_2$.
We can use this fact by optimising the contours for $f_1$ and $f_2$ separately.

Suppose on the other hand that $f(z) = g_1(z,z^\ast) + g_2(z,z^\ast)$, where $g_1(z,z^\ast)$ and $g_2(z,z^\ast)$
are now -- taken individually -- non-meromorphic functions.
If we now divide the integral into two channels, 
\bq
 I & = &
 \int\limits_{{\cal C}} dz \; g_1(z,z^\ast)
 +
 \int\limits_{{\cal C}} dz \; g_2(z,z^\ast),
\eq
we have to use the same contour for both channels.
However, we may use different parametrisations of the same contour.
This is the typical situation when we set $g_1(z,z^\ast) = w_1(z,z^\ast) f(z)$, 
$g_2(z,z^\ast) = w_2(z,z^\ast) f(z)$ with weight functions $w_1(z,z^\ast)$ and $w_2(z,z^\ast)$, satisfying $w_1+w_2=1$ and involving the
complex modulus.
This will be discussed in more detail in section \ref{sect:non_holomorphic}.

Let us now return to the integral in eq.~(\ref{complex_integral}):
\bq
\label{integral_starting_point}
 I
 & = & 
 \int\frac{d^{4}k}{(2\pi)^{4}}
 \frac{R(k)}{\prod\limits_{j=0}^{n-2} \left( k_j^2 - m_j^2 \right)}.
\eq
We define
\bq
 f_{\mathrm{UV}}\left(k\right)
 & = & 
 \prod\limits_{j=0}^{n-2} \frac{k_j^2 - m_j^2}{\bar{k}^2 - \mu_{\mathrm{UV}}^2}.
\eq
Clearly, $f_{\mathrm{UV}}(k)$ is a meromorphic function of $k$, with poles only at $\bar{k}^2-\mu_{\mathrm{UV}}^2=0$.
We can use the function $f_{\mathrm{UV}}(k)$ to split our original integral $I$ into two channels
\bq
\label{splitting_int_ext}
 I & = & I_{\mathrm{ext}} + I_{\mathrm{int}},
\eq
with
\bq
 I_{\mathrm{ext}} & = & \int \frac{d^4k}{(2\pi)^4} f_{\mathrm{UV}}\left(k\right) 
 \frac{R(k)}{\prod\limits_{j=0}^{n-2} \left( k_j^2 - m_j^2 \right)},
 \nonumber \\
 I_{\mathrm{int}} & = & \int \frac{d^4k}{(2\pi)^4} \left[ 1 - f_{\mathrm{UV}}\left(k\right) \right] 
 \frac{R(k)}{\prod\limits_{j=0}^{n-2} \left( k_j^2 - m_j^2 \right)}.
\eq
This splitting is holomorphic in $k$, therefore we can evaluate 
$I_{\mathrm{ext}}$ and $I_{\mathrm{int}}$ with two different contours.
The integrand of $I_{\mathrm{ext}}$ has only poles at $\bar{k}^2 - \mu_{\mathrm{UV}}^2 = 0$ and can be evaluated
with a relatively simple contour.
Since $\bar{k}=k-Q$ we further have within $I_{\mathrm{int}}$
\bq
 1 - f_{\mathrm{UV}}\left(k\right)
 & = &
 \frac{1}{\left(\bar{k}^2 - \mu_{\mathrm{UV}}^2\right)^{n-1}}
 \left\{ \left(k^2\right)^{n-2} 2 k \cdot \left[ \left( \sum\limits_{j=0}^{n-2} q_j \right) - \left(n-1\right) Q \right]
 + {\cal O}\left(k^{2n-4}\right) \right\}.
\eq
If we define the arbitrary four-vector $Q$ to be
\bq
\label{def_Q}
 Q & = & \frac{1}{(n-1)} \sum\limits_{j=0}^{n-2} q_j,
\eq
we can arrange that 
the integrand of $I_{\mathrm{int}}$ drops off with two extra powers
of $k$ for $|k| \rightarrow \infty$.
This ensures that $I_{\mathrm{int}}$ receives less contributions from the ultraviolet region.

In summary, the division into the two channels of eq.~(\ref{splitting_int_ext})
has the advantage that the integral $I_{\mathrm{ext}}$ has a simple pole structure, while the integral
$I_{\mathrm{int}}$ drops off with two additional powers in the ultraviolet region.

\subsection{Improvement of the ultraviolet subtraction terms}
\label{sect:uv_improvement}

Ultraviolet finiteness requires that the integrand falls of stronger than 
$1/|k|^4$ for $|k|\rightarrow \infty$.
The ultraviolet subtraction terms from ref.~\cite{Becker:2010ng} ensure that the integrand falls
of like
$1/|k|^5$ for $|k|\rightarrow \infty$.
This ensures that the integral is ultraviolet finite.
However, it turns out that the ultraviolet (or exterior) 
region contributes significantly to the statistical error.
In an expansion around the ultraviolet propagator $(\bar{k}^2-\mu_{\mathrm{UV}}^2)^{-1}$
a term which falls off like $1/|k|^5$ for $|k|\rightarrow \infty$ is given by
\bq
 \int \frac{d^4k}{(2\pi)^4} \frac{\bar{k} \cdot X}{\left(\bar{k}^2-\mu_{\mathrm{UV}}^2\right)^3},
\eq
where $X$ is a four-vector independent of $k$.
This term integrates to zero.
At the integrand level this term is oscillating. The integrand changes the sign under 
$\bar{k} \rightarrow (-\bar{k})$.
Oscillating terms pose no problem for a Monte Carlo integration, if the amplitude of the
oscillation is small compared to the terms which give a non-vanishing contribution after integration.
Unfortunately it turns out that the amplitude of the oscillations related to ultraviolet terms
of order $1/|k|^5$ or $1/|k|^6$ is not small. The amplitude is enhanced whenever an external
invariant approaches the jet resolution parameter.

The situation can be improved by a better damping in the ultraviolet region.
To this aim we modify the ultraviolet subtraction terms such that they subtract out also
the $1/|k|^5$- and the $1/|k|^6$-behaviour of the integrand, whenever there is a possibility
of having a small two-particle external invariant.
This amounts to modifying the ultraviolet subtraction terms for the propagators and the three-valent
vertices, as well as the soft and collinear subtraction terms.

Note that if one would like to achieve that the complete integrand has a behaviour better
than $1/|k|^6$ in the ultraviolet region, then it is not sufficient to include to the modifications
mentioned above the corresponding one for the four-gluon vertex.
In addition one has to include at ${\cal O}(|k|^{-6})$
subtraction terms for new four-, five- and six-valent vertices.
\begin{figure}
\begin{center}
\includegraphics{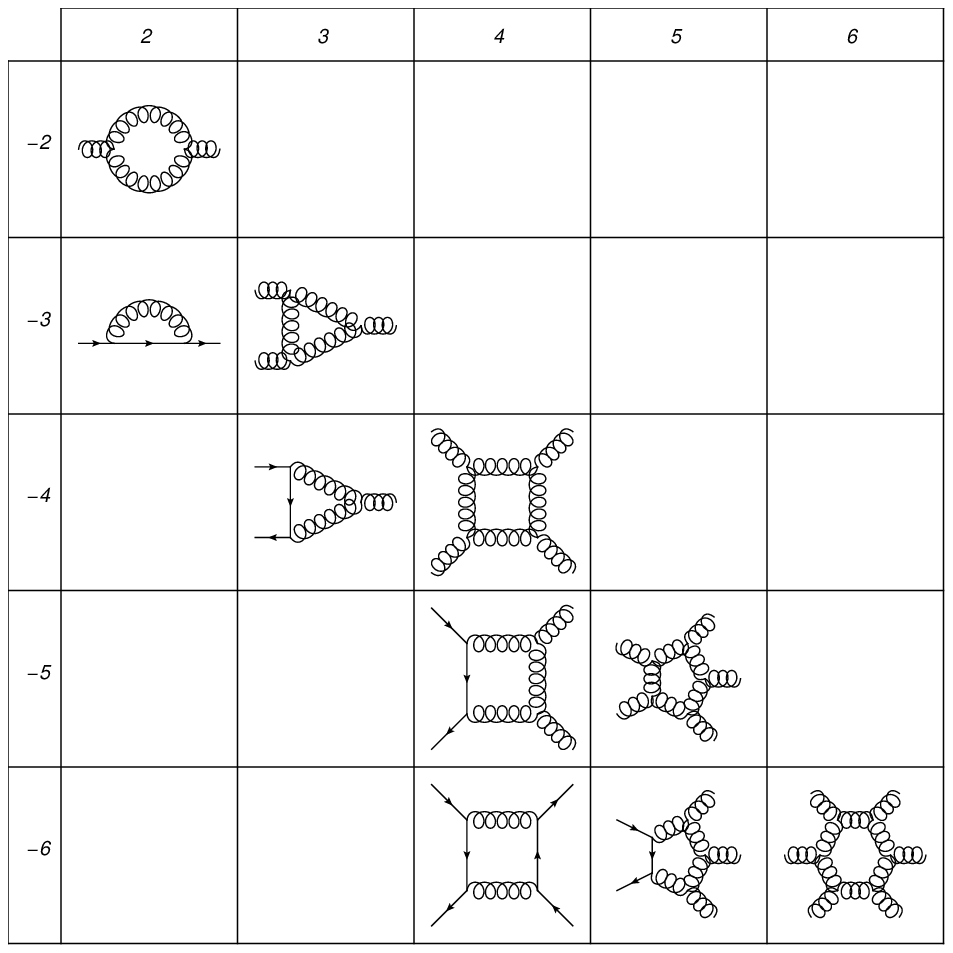}
\caption{\label{figure_uvtable}
The ultraviolet behaviour of some example diagrams. To the right the number of external particles
increases, in the vertical we have the various powers of the large $|k|$-behaviour, ranging
from $|k|^{-2}$ (top) to $|k|^{-6}$ (bottom).
}
\end{center}
\end{figure}
This is illustrated in fig.~(\ref{figure_uvtable}), where we show the ultraviolet behaviour for some
example diagrams.
From a technical point of view the inclusion of new subtraction terms for four-, five- and
six-valent vertices is not a problem, but the inclusion would modify the recursion relations for the
ultraviolet subtraction terms.
There is a trade off between the level of improvement in the subtraction terms and the efficiency for the
numerical evaluation of the integrand.
We have chosen to improve only the two- and three-valent parts of the ultraviolet behaviour.
Empirically it turns out that this is sufficient to reduce the oscillating behaviour in the 
ultraviolet region.

Technically the improvement is done as follows:
We have to improve eq.~(\ref{prop_uv_expansion}) to order
$|\bar{k}|^{-7}$.
This follows easily from
\bq
 \frac{1}{\left(k-p\right)^2-m^2}
  & = & 
 \frac{1}{\bar{k}^2-\mu_{\mathrm{UV}}^2}
 \left[ 1 - \frac{2 \bar{k} \cdot \left(p-Q\right)}{\bar{k}^2-\mu_{\mathrm{UV}}}
          + \frac{\left(p-Q\right)^2-m^2+\mu_{\mathrm{UV}}^2}{\bar{k}^2-\mu_{\mathrm{UV}}}
 \right]^{-1}
\eq
and the expansion
\bq
 \left[ 1 - a x + b x^2 \right]^{-1}
 & = & 1 + a x + \left(a^2 -b \right) x^2
 + \left( a^3 - 2 a b \right) x^3
 + \left( a^4 - 3 a^2 b + b^2 \right) x^4 
 \nonumber \\
 & &
 + {\cal O}\left(x^5\right),
\eq
with
\bq
 a = \frac{2 \bar{k} \cdot \left(p-Q\right)}{\bar{k}^2-\mu_{\mathrm{UV}}},
 & &
 b = \frac{\left(p-Q\right)^2-m^2+\mu_{\mathrm{UV}}^2}{\bar{k}^2-\mu_{\mathrm{UV}}}.
\eq
We apply this expansion to the propagator and the three-valent vertex corrections and determine
the ultraviolet behaviour up to order $|k|^{-6}$.
We then add appropriate terms of order $|k|^{-8}$ (which are beyond the order to which we are working)
to ensure that the integral over the ultraviolet subtraction terms is not changed.
In other words, we modify only the unintegrated version of the ultraviolet subtraction terms in such a way
that the integrated version of the ultraviolet subtraction terms remains unchanged.
Leaving the integrated version of the ultraviolet subtraction terms unchanged ensures that
the sum of all ultraviolet subtraction terms is proportional to a Born amplitude.
The explicit forms of the improved ultraviolet subtraction terms are rather lengthy.
We therefore do not list them here, however they can be obtained systematically with the 
procedure outlined above.

In addition we improve the ultraviolet behaviour of the soft and collinear subtraction terms, such that
they fall off like ${\cal O}(|k|^{-7})$ in the UV-region.
We give here the improved subtraction terms for massless QCD, the extension to massive quarks 
is straightforward.
The improved soft subtraction term is given by
\bq
\label{improved_soft}
 G_{\mathrm{soft}}^{(1)} & = &
 i
 \sum\limits_{j \in I_g}
 4 p_j \cdot p_{j+1}
 \left[ \frac{1}{k_{j-1}^2 k_j^2 k_{j+1}^2}  - \frac{1}{\left(\bar{k}^2-\mu_{\mathrm{UV}}^2\right)^3}  \right]
 A^{(0)}_j.
\eq
The second term in the square bracket is new and subtracts off the leading $|k|^{-6}$-behaviour.
Note that an individual soft subtraction term is proportional to a Born amplitude.
Therefore we can easily allow for a modification of the integrated soft subtraction term.
Integrating the soft subtraction term in eq.~(\ref{improved_soft})
we obtain
\bq
 S_\eps^{-1} \mu^{2\eps} \int \frac{d^Dk}{(2\pi)^D} G_{\mathrm{soft}}^{(1)} 
 & = &
 \mysign 
 \frac{1}{(4\pi)^2} 
 \frac{e^{\eps \gamma_E}}{\Gamma(1-\eps)}
 \sum\limits_{j \in I_g}
 \left[ 
 \frac{2}{\eps^2} \left( \frac{-2p_j\cdot p_{j+1}}{\mu^2} \right)^{-\eps}
 + \frac{2p_j\cdot p_{j+1}}{\mu_{\mathrm{UV}}^2}
 \right]
 A^{(0)}_j
 \nonumber \\
 & &
 + {\cal O}(\eps).
 \nonumber \\
\eq
The improved collinear subtraction is given by
\bq
 G_{\mathrm{coll}}^{(1)} & = & 
 \mynosign
 i
 \sum\limits_{j\in I_g}
 (-2) 
 \left( 
 \frac{S_j g_{\mathrm{UV}}\left(k_{j-1}^2,k_j^2\right) }{k_{j-1}^2 k_j^2}
+
 \frac{S_{j+1} g_{\mathrm{UV}}\left(k_{j}^2,k_{j+1}^2\right) }{k_{j}^2 k_{j+1}^2}
 \right)
 A^{(0)}_j,
\eq
where the function $g_{\mathrm{UV}}$ reads
\bq
\label{def2_g_UV}
\lefteqn{
 g_{\mathrm{UV}}\left(k_{j-1}^2,k_j^2\right)
 = 
 1 - \frac{k_{j-1}^2 k_j^2}{\left[ (k-Q)^2-\mu_{\mathrm{UV}}^2 \right]^2}
 - \frac{k_{j-1}^2 k_j^2 \left( 2 \bar{k} \cdot \bar{q}_{j-1} + 2 \bar{k} \cdot \bar{q}_{j} \right)}{\left[ (k-Q)^2-\mu_{\mathrm{UV}}^2 \right]^3}
 } & & \\
 & &
 + \frac{k_{j-1}^2 k_j^2 \left( \bar{q}_{j-1}^2 + \bar{q}_{j}^2  + 2 \mu_{\mathrm{UV}}^2 \right) }{\left[ (k-Q)^2-\mu_{\mathrm{UV}}^2 \right]^3}
 - \frac{k_{j-1}^2 k_j^2
   \left[ \left( 2 \bar{k} \cdot \bar{q}_{j-1} \right)^2
        + \left( 2 \bar{k} \cdot \bar{q}_{j} \right)^2
        + \left( 2 \bar{k} \cdot \bar{q}_{j-1} \right) \left( 2 \bar{k} \cdot \bar{q}_{j} \right) \right]
        }{\left[ (k-Q)^2-\mu_{\mathrm{UV}}^2 \right]^4}.
 \nonumber 
\eq
In eq.~(\ref{def2_g_UV}) we have used the notation
\bq
 \bar{q}_{j-1} = q_{j-1} - Q,
 & &
 \bar{q}_{j} = q_{j} - Q.
\eq
Integrating the collinear subtraction term we obtain
\bq
 S_\eps^{-1} \mu^{2\eps} \int \frac{d^Dk}{(2\pi)^D} G_{\mathrm{coll}}^{(1)} 
 & = & 
 \mysign 
 \frac{1}{(4\pi)^2} 
 \frac{e^{\eps \gamma_E}}{\Gamma(1-\eps)}
 \sum\limits_{j \in I_g}
  \left( S_j + S_{j+1} \right)
 \left[
 \frac{2}{\eps} \left( \frac{\mu_{\mathrm{UV}}^2}{\mu^2} \right)^{-\eps} 
 + 2
 \right]
 A^{(0)}_j
 + {\cal O}(\eps).
 \nonumber \\
\eq
Let us summarise what we achieved so far: We have split the integral in eq.~(\ref{integral_starting_point})
into two sub-integrals $I_{\mathrm{ext}}$ and $I_{\mathrm{int}}$.
With the improved ultraviolet subtraction terms, the integrand of $I_{\mathrm{ext}}$ falls off like $|k|^{-7}$
for all parts corresponding to one-loop $n$-point functions for $n\le 3$.
The integrand of $I_{\mathrm{ext}}$ falls off like $|k|^{-5}$
for the parts corresponding to one-loop $n$-point functions for $n \ge 4$.
On the other hand, the integrand of $I_{\mathrm{int}}$ is suppressed by two additional powers
of $|k|$.
Therefore the parts corresponding to one-loop $n$-point functions with $n\le 3$ in 
$I_{\mathrm{int}}$ fall off like $|k|^{-9}$, while the 
parts corresponding to one-loop $n$-point functions with $n \ge 4$  
fall off like $|k|^{-7}$.

We can further improve the ultraviolet behaviour of all terms by another power of $|k|$ as follows:
All terms which scale with an odd power in the ultraviolet region are necessarily anti-symmetric under the
substitution $\bar{k} \rightarrow (-\bar{k})$.
We can eliminate these terms by sampling simultaneously the points $\bar{k}$ and $(-\bar{k})$.
With this method we can reduce the leading ultraviolet behaviour (which scales like an odd power in all cases above) 
to the next lower even power.
This will be discussed in more detail in sub-section~\ref{sect:sampling}.

\subsection{Optimisation of the integration contour}

Since the division of the integral in eq.~(\ref{integral_starting_point})
into the two sub-integrals $I_{\mathrm{ext}}$ and $I_{\mathrm{int}}$
is holomorphic, we may use different integration contours for $I_{\mathrm{ext}}$ and $I_{\mathrm{int}}$.

\subsubsection{The contour deformation for $I_{\mathrm{ext}}$}

By construction the integrand of $I_{\mathrm{ext}}$ has only poles at
$\bar{k}^2-\mu_{\mathrm{UV}}^2=0$.
The poles are located on a single light cone in loop momentum space and the contour deformation is rather simple.
For $I_{\mathrm{ext}}$ we use the contour deformation
\bq
\label{deformation_exterior}
 k & = & \tilde{k} + i \kappa
\eq
with
\bq
 \kappa^\mu & = & g_{\mu\nu} \left( \tilde{k}^\nu - Q^\nu \right).
\eq
(The position of the indices is correct. We use the metric tensor $g_{\mu\nu}=\mbox{diag}(1,-1,-1,-1)$ to indicate
that the spatial components of contravariant four-vector $\kappa^\mu$ are the negative of the spatial components
of the contravariant four-vector $\tilde{k}^\mu-Q^\mu$.)
We have
\bq
 \bar{k}^2-\mu_{\mathrm{UV}}^2
 & = &
 2 i \left( \tilde{k} - Q \right) \circ \left( \tilde{k} - Q \right) - \mu_{\mathrm{UV}}^2,
\eq
where $a \circ b$ denotes the Euclidean scalar product of the four-vectors $a$ and $b$.
As already mentioned we will always choose $\mu_{\mathrm{UV}}^2$ purely imaginary with
$\mbox{Im}\;\mu_{\mathrm{UV}}^2 < 0$.
Therefore the term $-\mu_{\mathrm{UV}}^2$ alone ensures that the imaginary part of $\bar{k}^2-\mu_{\mathrm{UV}}^2$ is always
positive.
The reader may ask why we deform the integration contour at all. The answer is given by the ultraviolet behaviour
of the propagator $[\bar{k}^2-\mu_{\mathrm{UV}}^2]^{-1}$: With the deformation as in eq.~(\ref{deformation_exterior})
the propagator falls off always like $|\tilde{k}|^{-2}$ for $|\tilde{k}|\rightarrow \infty$, while with no deformation the
propagator would remain constant along the light cone $\bar{k}^2=0$.
The Jacobian of the contour deformation is given by
\bq
 \left|\frac{\partial k^{\mu}}{\partial \tilde{k}^{\nu}}\right|
 & = &
 - 4 i.
\eq

\subsubsection{The contour deformation for $I_{\mathrm{int}}$}

The integrand of $I_{\mathrm{int}}$ has good ultraviolet properties, but it has a more complicated
infrared structure.
With $\mu_{\mathrm{UV}}^2$ purely imaginary and
$\mbox{Im}\;\mu_{\mathrm{UV}}^2 < 0$, we have to take into account the poles given by
\bq
 k_j^2 - m_j^2 & = & 0,
 \;\;\;\;\;\; 0 \le j \le n-2.
\eq
In this section we discuss the massless case $m_j=0$, so the poles are given by $k_j^2=0$ for $0 \le j \le (n-2)$.
For $I_{\mathrm{int}}$ we use the contour deformation along the lines of ref.~\cite{Gong:2008ww}. 
We first define
\bq
 P & = & \frac{1}{2} \left( q_0 + q_{n-2} \right).
\eq
The four-vector $P$ is the centre of the forward light cone from $q_0$ intersected with the
backward light cone from $q_{n-2}$.
We recall that the four-vector $Q$ has been defined in eq.~(\ref{def_Q}) as the average of the $(n-1)$ four-vectors $q_0$, ..., $q_{n-2}$.
Note that in general $P$ and $Q$ are not equal.
We further define
\bq
 x = -\frac{2 \tilde{k}_{n-2} \cdot p_{n-1}}{\left( q_{n-2} - q_0 \right)^2},
 & &
 \bar{x} = -\frac{2 \tilde{k}_{n-1} \cdot p_{n}}{\left( q_{n-2} - q_0 \right)^2}.
\eq
We set
\bq
\label{def_gns_contour}
 k = \tilde{k} + i \lambda \kappa,
 & &
 \kappa^\mu =  
 \left( c_+ - c_- \right) \left( q_{n-2} - q_0 \right) - \sum\limits_{j=0}^{n-2} c_j \tilde{k}_j.
\eq
The coefficients $c_\pm$ and $c_j$ ($0 \le j \le n-2$) are given by
\bq
 c_+ & = & \left( x + \bar{x} \right) \theta\left( x + \bar{x} \right)  h_-\left(\tilde{k}_{n-2}\right) g_-\left(\tilde{k}-P\right),
 \nonumber \\
 c_- & = & \left( -x - \bar{x} \right) \theta\left( -x - \bar{x} \right)  h_+\left(\tilde{k}_{0}\right) g_+\left(\tilde{k}-P\right),
 \nonumber \\
 c_0 & = & h_+\left(\tilde{k}_{1}\right) g\left(\tilde{k}-P\right),
 \nonumber \\
 c_j & = & h_+\left(\tilde{k}_{j+1}\right) h_-\left(\tilde{k}_{j-1}\right) g\left(\tilde{k}-P\right),
 \;\;\; 1 \le j \le n-3,
 \nonumber \\
 c_{n-2} & = & h_-\left(\tilde{k}_{n-3}\right) g\left(\tilde{k}-P\right).
\eq
The functions $h_+$, $h_-$, $g_+$, $g_-$ and $g$ are defined in \cite{Gong:2008ww} and read
\bq
 h_+\left(k\right) =  
  \frac{\left( \left|\vec{k}\right| - k^0 \right)^2}{\left( \left|\vec{k}\right| - k^0 \right)^2 + M_1^2} 
  \theta\left( \left|\vec{k}\right| - k^0 \right),
 & &
 h_-\left(k\right) =  
  \frac{\left( \left|\vec{k}\right| + k^0 \right)^2}{\left( \left|\vec{k}\right| + k^0 \right)^2 + M_1^2} 
  \theta\left( \left|\vec{k}\right| + k^0 \right),
\eq
\bq
 g\left(k\right) = \frac{\gamma_1 M_2^2}{k \circ k + M_2^2},
 & &
 g_\pm\left(k\right) = \frac{\gamma_2}{1+\left(1 \pm \frac{k^0}{\sqrt{\left| \vec{k} \right|^2 + M_3^2}}\right)^2}.
\eq
These functions depend on the five parameters $M_1$, $M_2$, $M_3$, $\gamma_1$ and $\gamma_2$.  
The default choice for these parameters is
\bq
M_1 = 0.05 \sqrt{\frac{1}{2}\left( q_{n-2} - q_0 \right)^2},
 & &
M_2 = M_3 = \sqrt{\frac{1}{2}\left( q_{n-2} - q_0 \right)^2},
\eq
\bq
 \gamma_1 = 0.7,
 & & 
 \gamma_2 = 1.
\eq
These default parameters have been given in ref.~\cite{Gong:2008ww}.
We have varied these parameters in different processes. It turns out that these values are a good choice for a wide range of processes.
It remains to define the scaling parameter $\lambda$.
We set 
\bq
 \lambda & = & \min\left( 1, \lambda_0, ..., \lambda_{n-2}, \lambda_{\mathrm{UV}}, \lambda_{\mathrm{coll}} \right),
\eq
with $\lambda_j$ ($0 \le j \le n-2$) given by
\bq
 \lambda_j & = & 
 \left\{ \begin{array}{lll}
 \left( \frac{\kappa^2 \tilde{k}_j^2}{\left(2 \kappa^2 \right)^2} \right)^{\frac{1}{2}} & \mbox{for} & 0 < 2 \left( \kappa \cdot \tilde{k}_j \right)^2 < \kappa^2 \tilde{k}_j^2, \\
 \left( \frac{4 \left( \kappa \cdot \tilde{k}_j \right)^2 - \kappa^2 \tilde{k}_j^2}{\left(2 \kappa^2 \right)^2} \right)^{\frac{1}{2}}  & \mbox{for} & 0 < \kappa^2 \tilde{k}_j^2 < 2 \left( \kappa \cdot \tilde{k}_j \right)^2, \\
 \left( \frac{4 \left( \kappa \cdot \tilde{k}_j \right)^2 - 2 \kappa^2 \tilde{k}_j^2}{\left(2 \kappa^2 \right)^2} \right)^{\frac{1}{2}} & \mbox{for} & \kappa^2 \tilde{k}_j^2 < 0 < 2 \left( \kappa \cdot \tilde{k}_j \right)^2. \\
 \end{array} \right.
\eq
$\lambda_{\mathrm{coll}}$ is given by
\bq
 \lambda_{\mathrm{coll}} & = & \frac{1}{4C},
 \;\;\;
 C = \sum\limits_{j=0}^{n-2} c_j.
\eq
Finally, $\lambda_{\mathrm{UV}}$ is given by
\bq
 \lambda_{\mathrm{UV}} & = &
 \left\{ \begin{array}{ll}
 1 & \mbox{for}\; 4 \left( \tilde{k} - Q \right) \cdot \kappa > \mbox{Im} \mu_{\mathrm{UV}}^2\\
 \frac{\mbox{Im} \mu_{\mathrm{UV}}^2}{4 \left( \tilde{k} - Q \right)\cdot \kappa} & \mbox{otherwise}.\\
 \end{array} \right.
\eq
$\lambda_0$, ..., $\lambda_{n-2}$ and $\lambda_{\mathrm{coll}}$ are defined as in ref.~\cite{Gong:2008ww}.
In addition we introduced $\lambda_{\mathrm{UV}}$, which protects the ultraviolet propagator $\bar{k}^2-\mu_{\mathrm{UV}}^2$ from going
on-shell due to a too large contour deformation.
Note that $\lambda_{\mathrm{UV}}$ differs from $1$ only for 
\bq
 4 \left( \tilde{k} - Q \right) \cdot \kappa < \mbox{Im} \mu_{\mathrm{UV}}^2 < 0.
\eq
It can be shown that with the contour deformation as in eq.~(\ref{def_gns_contour})
the propagators fall off like $|\tilde{k}|^{-2}$ for $|\tilde{k}|\rightarrow \infty$.
For the proof it is convenient to discuss the time-like, space-like and light-like regions separately.
In the time-like ($\tilde{k}^0\rightarrow \infty$, $\vec{\tilde{k}}=\mbox{const}$) and space-like ($|\vec{\tilde{k}}|\rightarrow \infty$, $\tilde{k}^0=\mbox{const}$)
regions the contour deformation goes to zero in the limit of $|\tilde{k}|\rightarrow \infty$.
Therefore $k^2 \approx \tilde{k}^2$ and the large $|\tilde{k}|$-behaviour is given by the real part.
In the light-like region ($\tilde{k}^2\approx 0$) the contour deformation does not vanish in the 
limit where $|\tilde{k}|\rightarrow \infty$ and the imaginary part of $k^2$ scales like $|\tilde{k}|^2$.
\\
The Jacobian 
\bq
 \left|\frac{\partial k^{\mu}}{\partial \tilde{k}^{\nu}}\right|
\eq
for the contour deformation for $I_{\mathrm{int}}$ is computed numerically.

\subsection{Sampling in loop momentum space}
\label{sect:sampling}

After the contour deformation we have a four-dimensional integral
\bq
 \int \frac{d^4\tilde{k}}{(2\pi)^4} f(\tilde{k})
\eq
over the four real variables $\tilde{k}^0$, $\tilde{k}^1$, $\tilde{k}^2$ and $\tilde{k}^3$.
For each dimension the integration is from minus infinity to plus infinity.
In order to be able to use standard methods for the Monte Carlo integration like Vegas \cite{Lepage:1978sw,Lepage:1980dq}
we map ${\mathbb R}^4$ to the four-dimensional unit hyper-cube $[0,1]^4$.
There are many possible choices for such a mapping.
We choose a mapping which approximates the peak structure of the integrand.
This amounts to importance sampling and reduces the statistical Monte Carlo error.
In order to further minimise the Monte Carlo integration error
we use where possible the method of antithetic variates.
Within this method we always evaluate a few points together, which are anti-correlated.
In this way oscillations are significantly reduced.

\subsubsection{Sampling for $I_{\mathrm{ext}}$}

We start the discussion with our method for sampling the integrand of $I_{\mathrm{ext}}$.
We generate the real vector $\bar{k}_{\mathrm{real}}=\tilde{k}-Q$ 
as follows:
We use four uniformly in $[0,1]$ distributed random numbers $u_0$, ..., $u_3$ and
define the four quantities $k_E$, $\xi$, $\theta$ and $\phi$ by the equations
\bq
 k_E & = & \mu_1 \sqrt{ \tan \frac{\pi}{2} u_0 },
 \nonumber \\
 \xi & = & \arccos\left(1-2u_1\right),
 \nonumber \\
 \theta & = & \arccos\left(1-2u_2\right),
 \nonumber \\
 \phi & = & 2 \pi u_3.
\eq
$\mu_1$ is an arbitrary scale, which we take to be of the order of the centre-of-mass energy.
We then set
\bq
 \bar{k}^0_{\mathrm{real}} = k_E \cos \xi,
 \;\;\;
 k_r = k_E \sin \xi,
 \;\;\;
 \vec{\bar{k}}_{\mathrm{real}} = k_r \left( \begin{array}{c}
 \sin \theta \sin \phi \\
 \sin \theta \cos \phi \\
 \cos \theta
 \end{array} \right).
\eq
The Jacobian of this transformation
is
\bq
 \left| \frac{\partial \tilde{k}}{\partial u} \right| 
 = 2 \pi^2 \frac{k_E^2}{\mu_1^2} \left( k_E^4 + \mu_1^4 \right) \sin \xi.
\eq
In section~\ref{sect:uv_improvement} we have shown that the integrand of $I_{\mathrm{ext}}$ falls off like
$|k|^{-7}$ for all parts corresponding to one-loop $n$-point functions for $n\le 3$, and like
$|k|^{-5}$ for the parts corresponding to one-loop $n$-point functions for $n \ge 4$.
By sampling always the two points with loop momenta $\bar{k}_{\mathrm{real}}$ and $(-\bar{k}_{\mathrm{real}})$ together we can 
reduce the ultraviolet behaviour to $|k|^{-8}$ and $|k|^{-6}$, respectively.

\subsubsection{Division into sub-channels for $I_{\mathrm{int}}$}
\label{sect:non_holomorphic}

The efficient sampling of the integrand of $I_{\mathrm{int}}$ is more involved.
We recall from fig.~(\ref{fig_zig_zag}) that the integrand is characterised by a strand 
of $(n-2)$ line segments (diagram $(c)$ of fig.~(\ref{fig_zig_zag})).
We view a single line segment as a basic building block 
and we divide $I_{\mathrm{int}}$ into $(n-2)$ sub-channels, such that each sub-channel corresponds to a line segment.
\begin{figure}
\begin{center}
\includegraphics[bb=90 600 515 735,width=0.9\textwidth]{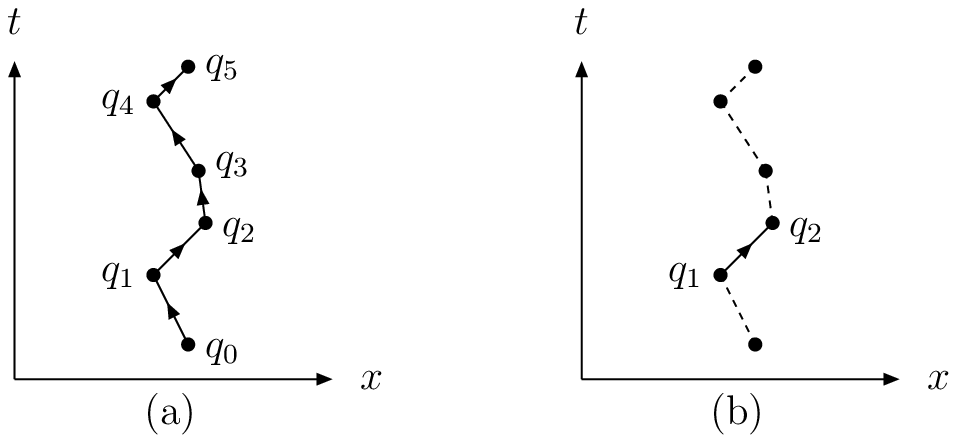}
\caption{\label{fig_zig_zag2}
Diagram $(a)$ shows the origins of the light cones in electron-positron annihilation. 
The origins are given by the $(n-1)$ vertices. The vertices are connected by $(n-2)$ line segments.
We decompose $I_{\mathrm{int}}$ into $(n-2)$ sub-channels, such that each sub-channel corresponds to one line segment.
This is shown in diagram $(b)$, where the line segment from $q_1$ to $q_2$ is drawn.
}
\end{center}
\end{figure}
This is shown pictorially in fig.~(\ref{fig_zig_zag2}).
Technically this is done as follows:
After contour deformation we have an integral over a real four-dimensional space
\bq
 I_{\mathrm{int}} & = & \int \frac{d^4\tilde{k}}{(2\pi)^4} f\left(\tilde{k}\right).
\eq
The line segments correspond to 
\bq
 \tilde{k} & = & q_j + x \left( q_{j+1} - q_j \right),
 \;\;\; 0 \le x \le 1,
 \;\;\; 0 \le j \le n-3
\eq
in loop momentum space.
We have $q_{j+1}-q_j=p_{j+1}$. We split the original integral into several channels, such that
each critical line segment corresponds to a separate channel.
We then use for each channel a dedicated mapping for the loop momentum.
The splitting into different channels is done as follows: We re-write the original integral
as
\bq
 I_{\mathrm{int}} & = & \sum\limits_{i=0}^{n-3} \int \frac{d^4\tilde{k}}{(2\pi)^4} w_i\left(\tilde{k}\right) f\left(\tilde{k}\right),
\eq
with
\bq
 w_i \ge 0 & \mbox{and} &
 \sum\limits_{i=0}^{n-3} w_i\left(\tilde{k}\right) = 1.
\eq
Since the sum over all weights $w_i$ equals one, the sum over all channels equals the original integral.
For the weights $w_i$ we use
\bq
 w_i\left(\tilde{k}\right) & = & \frac{ \left( \frac{1}{ \left| k_i^2 \right| \left| k_{i+1}^2 \right| } \right)^\alpha}
                { \sum\limits_{j=0}^{n-3} \left( \frac{1}{ \left| k_j^2 \right| \left| k_{j+1}^2 \right| } \right)^\alpha},
\eq
with $\alpha=2$.
Note that for the weights we take the norm of the complex quantities $k_j^2$.
Therefore the division into the sub-channels is not holomorphic and we have to use the same deformation contour for all
sub-channels.
The weights have the properties that
\bq
 \lim \; w_i=1 & \mbox{if} & 
 \tilde{k} \rightarrow q_i + x \left( q_{i+1} - q_i \right)
\eq
and
\bq
 \lim \; w_i=0 & \mbox{if} &
 \tilde{k} \rightarrow q_j + x \left( q_{j+1} - q_j \right) \;\; \mbox{with} \;\; i \neq j.
\eq
This ensures that in each channel there is only one critical line segment.

\subsubsection{Sampling of an individual sub-channel of $I_{\mathrm{int}}$}

After having divided $I_{\mathrm{int}}$ into different sub-channels we can discuss how to sample the integration points for
a specific sub-channel.
Recall that a line segment in loop momentum space goes from $q_j$ to $q_{j+1}$.
The four-vector from $q_j$ to $q_{j+1}$ is given by
$p_{j+1}=q_{j+1}-q_j$. 
Since we are discussing a single line segment we will in the following simply write $p$ instead of $p_{j+1}$.
For massless external particles $p$ is a light-like four-vector.
We now look for an appropriate coordinate system in the case when there is a distinguished vector $p$.
A possible choice are generalisations of elliptical or prolate spheroidal coordinate systems to four dimensions.

In detail we generate $\tilde{k}$ as follows:
Let $p$ be a light-like four-vector. In spherical coordinates $p$ can be written as
\bq
 p^0 & = & \left|p\right| \cos \theta_1,
 \nonumber \\
 p^1 & = & \left|p\right| \sin \theta_1 \cos \theta_2,
 \nonumber \\
 p^2 & = & \left|p\right| \sin \theta_1 \sin \theta_2 \cos \phi_3,
 \nonumber \\
 p^3 & = & \left|p\right| \sin \theta_1 \sin \theta_2 \sin \phi_3,
\eq
where
\bq
 \left|p\right| = \sqrt{ \sum\limits_{i=0}^3 \left(p^i\right)^2},
 \;\;\;
 \theta_1 = \arccos \frac{p^0}{\left|p\right|},
 \;\;\;
 \theta_2 = \arctan \frac{\sqrt{ \left(p^2\right)^2 + \left(p^3\right)^2}}{p^1},
 \;\;\;
 \phi_3 = \arctan \frac{p^3}{p^2}.
\eq
The angles $\theta_1$, $\theta_2$ and $\phi_3$ define three rotation matrices
\bq
 R_1 = \left(\begin{array}{cccc}
 \cos \theta_1 & - \sin \theta_1 & 0 & 0 \\
 \sin \theta_1 & \cos \theta_1 & 0 & 0 \\
 0 & 0 & 1 & 0 \\
 0 & 0 & 0 & 1 \\
\end{array} \right),
 \;\;\;
 R_2 = \left(\begin{array}{cccc}
 1 & 0 & 0 & 0 \\
 0 & \cos \theta_2 & - \sin \theta_2 & 0 \\
 0 & \sin \theta_2 & \cos \theta_2 & 0 \\
 0 & 0 & 0 & 1 \\
\end{array} \right),
\eq
\bq
 R_3 = \left(\begin{array}{cccc}
 1 & 0 & 0 & 0 \\
 0 & 1 & 0 & 0 \\
 0 & 0 & \cos \phi_3 & - \sin \phi_3 \\
 0 & 0 & \sin \phi_3 & \cos \phi_3 \\
\end{array} \right).
\eq
We generate the (real) loop momentum $\tilde{k}$ as follows:
\bq
\label{def_loop_momentum_prolate}
 \tilde{k} & = & q_j + \frac{1}{2} p + R_3 \cdot R_2 \cdot R_1 \cdot \tilde{k}',
\eq
where $p=p_{j+1}$ and 
\bq
 \tilde{k}' & = &
 \frac{1}{2} \left|p\right|
 \left( \begin{array}{l}
  \cosh \rho \; \cos \xi \\
  \sinh \rho \; \sin \xi \; \cos \theta \\
  \sinh \rho \; \sin \xi \; \sin \theta \; \cos \phi \\
  \sinh \rho \; \sin \xi \; \sin \theta \; \sin \phi \\
 \end{array} \right).
\eq
$\rho$, $\xi$, $\theta$ and $\phi$ are generalised elliptical coordinates.
The coordinate $\rho$ ranges from $[0,\infty[$, the coordinates $\xi$ and $\theta$ have the range
$[0,\pi]$, whereas $\phi$ is in the range $[0,2\pi]$.
The Jacobian is given by
\bq
 \left| \frac{\partial \tilde{k}}{\partial \tilde{k}'} \right|
 & = &
 \frac{1}{16} \left| p \right|^4 
 \sinh^2 \rho \; \sin^2 \xi \; \sin \theta \; \left( \sinh^2 \rho + \sin^2 \xi \right).
\eq
The variables $(\rho,\xi,\theta,\phi)$ we generate as follows:
\bq
\label{functions_prolate}
 \rho & = & \ln\left( 1 + \frac{\mu_0}{\left|p\right|} \tan \frac{\pi}{2} u_0 \right),
 \nonumber \\
 \xi & = & \pi u_1,
 \nonumber \\
 \theta & = & \left\{ \begin{array}{ll}  
                     \arccos\left[ \left(1+\eps\right) \left(\frac{1+\eps}{\eps}\right)^{-2 u_2 } - \eps\right] & 0 \le u_2 < \frac{1}{2}, \\
                     \arccos\left[ \eps - \left(1+\eps\right) \left(\frac{1+\eps}{\eps}\right)^{-2 (1-u_2) } \right] & \frac{1}{2} \le u_2 < 1, \\
              \end{array} \right.
 \nonumber \\
 \phi & = & 2 \pi u_3,
\eq
with
\bq
 \eps & = & \sinh \rho \; \sin \xi.
\eq
$\mu_0$ is again an arbitrary scale, which we also take to be of the order of the centre-of-mass energy.
The functions in eq.~(\ref{functions_prolate}) have been chosen such that they approximate the typical peak structure
of the integrand.
The Jacobians are
\bq
 \frac{\partial \rho}{\partial u_0} & = & \frac{\pi}{2} \frac{\left[\frac{\mu_0^2}{\left|p\right|^2} + \left( e^\rho - 1 \right)^2\right]}{\frac{\mu_0}{\left|p\right|} e^\rho},
 \nonumber \\
 \frac{\partial \xi}{\partial u_1} & = & \pi,
 \nonumber \\
 \frac{\partial \theta}{\partial u_2} & = & 2 \frac{\left(\eps+\left|\cos\theta\right|\right)}{\sin \theta} \ln\left(\frac{1+\eps}{\eps}\right),
 \nonumber \\
 \frac{\partial \phi}{\partial u_3} & = & 2 \pi.
\eq
Again we use the method of antithetic variates: 
We observe that the integrand has a periodic behaviour in $\phi$. 
Therefore combining the evaluations at $\phi$ and $(\phi+\pi) \; \mbox{mod} \; (2\pi)$ averages out these
oscillations.
In addition, the integrand is for $\rho \rightarrow 0$ strongly peaked and antisymmetric around $\theta=\pi/2$.
Evaluating the integrand at $\theta$ and $\pi-\theta$ averages out this behaviour.
Furthermore we evaluate the integrand always 
with the values $\tilde{k}'$ and $(-\tilde{k}')$ in eq.~(\ref{def_loop_momentum_prolate}).
This improves the ultraviolet behaviour.

\section{Recurrence relations}
\label{sect:recurrence}

This section is devoted to the computation of the integrands.
The computation of the integrands is done efficiently with the help of recurrence relations.
We first discuss in sub-section \ref{sect:tree_recursion} tree-level recurrence relations, which are directly relevant to the Born contribution, the integrated
subtraction terms and the unintegrated soft and collinear subtraction terms.
In sub-section \ref{sect:loop_recursion} we discuss how the integrand of the bare one-loop amplitude is calculated.
Cutting the loop open reduces the computation to a tree-like problem.
In sub-section \ref{sect:uv_recursion} we treat the computation of the ultraviolet subtraction term.
Again, this is done recursively with a modified tree-like recurrence relation.

\subsection{Tree level recurrence relations}
\label{sect:tree_recursion}

We first review the computation of tree-level partial amplitudes.
Berends-Giele type recurrence relations \cite{Berends:1987me}
build tree-level partial amplitudes from smaller building blocks, 
called colour ordered off-shell currents.
Off-shell currents are objects with $n$ on-shell legs and one additional leg off-shell.
Momentum conservation is satisfied. It should be noted that
off-shell currents are not gauge invariant objects.
Recurrence relations relate off-shell currents with $n$ legs 
to off-shell currents with fewer legs.
\begin{figure}
\begin{center}
\includegraphics[bb=45 670 560 785,width=0.9\textwidth]{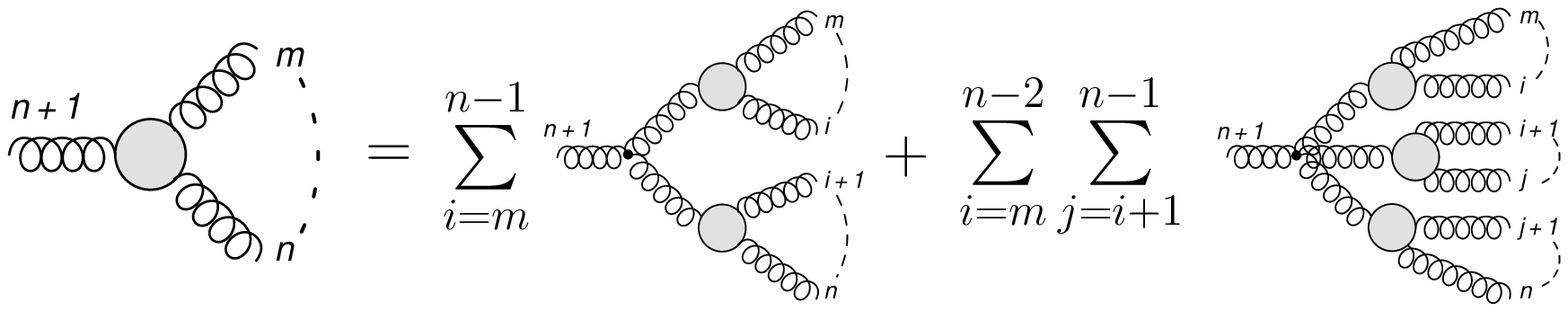}
\end{center}
\caption{\label{fig:treelevelgluon} The recurrence relation for the pure gluon current.}
\end{figure}
The recursion relation for the pure gluon off-shell current is depicted in fig.\ref{fig:treelevelgluon}.
The recursion starts with the one-currents:
\bq
J^{(0)}_\mu(l) & = & \eps_\mu^{\lambda}\left(p_l,q_l\right).
\eq
$\eps_\mu^\lambda$ is the polarisation vector of the gluon corresponding to the polarisation $\lambda$,
$p_l$ the four-momen\-tum of the gluon and $q_l$ an arbitrary light-like reference momentum used to define the polarisation of the 
gluon.
The recursive relation states that in the pure gluon off-shell current
a gluon couples to other gluons only via the three- or four-gluon
vertices:
\bq
\label{Berends_Giele_recursion}
 J^{(0)}_\alpha(m,...,n) & = & 
 \frac{-i g_{\alpha\mu}}{P^2_{m,n}} 
 \left[ 
        \sum\limits_{j=m}^{n-1} V_3^{\mu\nu\rho}(-P_{m,n},P_{m,j},P_{j+1,n})
                                J^{(0)}_\nu(m,...,j) J^{(0)}_\rho(j+1,...,n) 
 \right. \nonumber \\
 & & \left. 
        + \sum\limits_{j=m}^{n-2} \; \sum\limits_{l=j+1}^{n-1} V_4^{\mu\nu\rho\sigma} 
            J^{(0)}_\nu(m,...,j) J^{(0)}_\rho(j+1,...,l) J^{(0)}_\sigma(l+1,...,n) 
     \right], 
\;\;\;\;\;
\eq
where
\bq
P_{i,j} & = & p_i + p_{i+1} + ... + p_j 
\eq
and $V_3$ and $V_4$ are the colour ordered three-gluon and four-gluon vertices
\bq
\label{Feynman_rules}
 V_3^{\mu\nu\rho}(p_1,p_2,p_3) 
 & = & 
 i \left[
          g^{\mu\nu} \left( p_2^\rho - p_1^\rho \right)
        + g^{\nu\rho} \left( p_3^\mu - p_2^\mu \right)
        + g^{\rho\mu} \left( p_1^\nu - p_3^\nu \right)
   \right],
 \nonumber \\
 V_4^{\mu\nu\rho\sigma} & = & i \left( 2 g^{\mu\rho} g^{\nu\sigma} - g^{\mu\nu} g^{\rho\sigma} -g^{\mu\sigma} g^{\nu\rho} \right).
\eq
From an off-shell current one easily recovers the on-shell amplitude by 
removing the extra propagator,
taking the leg $(n+1)$ on-shell
and contracting with the appropriate polarisation vector:
\bq
A^{(0)}(1,...,n+1) 
 & = & 
 \eps^\mu(p_{n+1},q_{n+1}) i P_{1,n}^2 J^{(0)}_\mu(1,...,n) \big|_{P_{1,n}=-p_{n+1}}.
\eq
Similar recurrence relations can be written down for the quark and antiquark currents, as well as the gluon
currents in full QCD.
If there is only one quark line, we have for the quark and antiquark off-shell currents
\bq
\label{tree_off_shell_current_quark}
\overline{U}^{(0)}(m,...,n) 
 & = & 
  i \sum\limits_{i=m}^{n-1}\overline{U}^{(0)}(m,...,i) V_{qg\bar{q}}^{\mu}J_{\mu}(i+1,...,n)
  \frac{{P\sla}_{m,n}}{P_{m,n}^2},
 \nonumber \\
V^{(0)}(m,...,n) 
 & = &
 -i 
 \frac{{P\sla}_{m,n}}{P_{m,n}^2}
 \sum\limits_{i=m}^{n-1}V_{qg\bar{q}}^{\mu}J_{\mu}(m,...,i)V^{(0)}(i+1,...,n).
\eq
This is shown pictorially in fig.\ref{fig:treelevelquark}.
\begin{figure}
\begin{center}
\includegraphics[bb=110 655 460 785,width=0.6\textwidth]{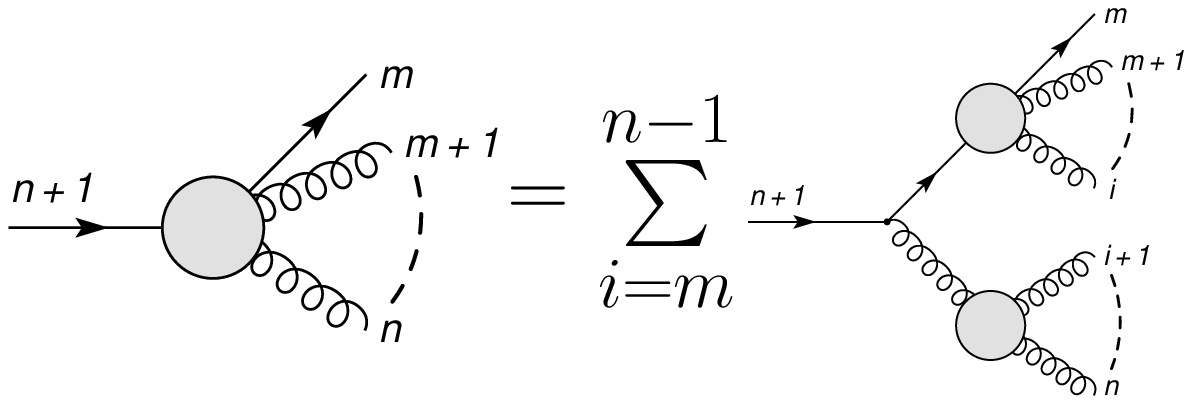}
\includegraphics[bb=110 655 460 785,width=0.6\textwidth]{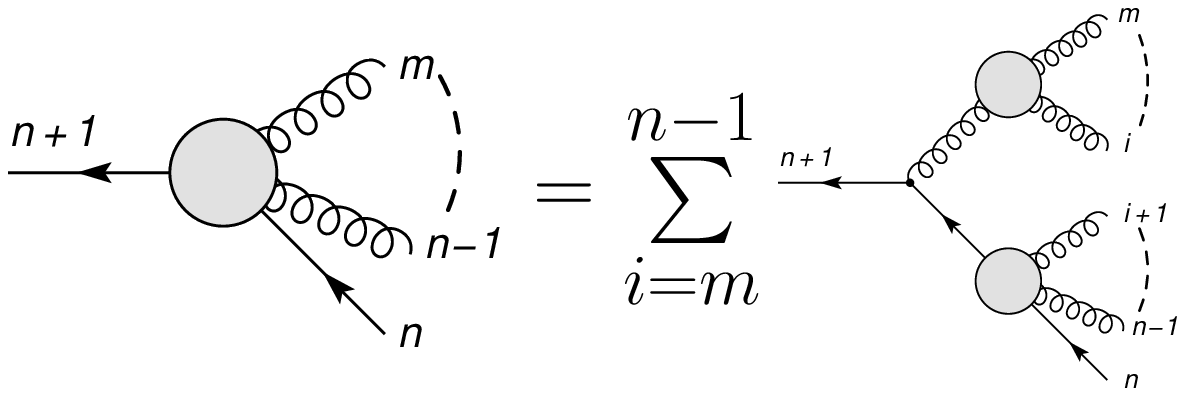}
\end{center}
\caption{\label{fig:treelevelquark} The recurrence relations for the quark current and the antiquark current.}
\end{figure}
The recursions start with $\overline{U}^{(0)}(l)=\bar{u}(p_{l})$ and $V^{(0)}(l)=v(p_{l})$, respectively. 
The colour ordered quark-gluon vertex is given by 
\bq
 V_{qg\bar{q}}^{\mu} & = & -i\gamma^\mu.
\eq
Note that in eq.~(\ref{tree_off_shell_current_quark}) the quantities
$\overline{U}^{(0)}$, $V^{(0)}$, $V_{qg\bar{q}}$ and $P\sla$ are matrices in Dirac space,
and the order is relevant.
The partial amplitudes with one quark-antiquark line are given by
\bq
A^{(0)}(1_q,2,...,n-1,n_{\bar{q}})
 & = &
 -i \overline{U}^{(0)}(1,2,...,n-1) {P\sla}_{1,n-1} v(p_{n}) \big|_{P_{1,n-1}=-p_{n}}
 \nonumber \\
 & = & 
 i \bar{u}(p_{1}) {P\sla}_{2,n} V^{(0)}(2,...,n-1,n) \big|_{P_{2,n}=-p_{1}}.
\eq
The quark and antiquark off-shell currents enter also the computation for the amplitudes
\bq
 e^+ e^- \rightarrow q, g, ..., g, \bar{q}.
\eq
Taking all particles as outgoing we have
\bq
\lefteqn{
A^{(0)}(1_q,2_g,...,(n-3)_g,(n-2)_{\bar{q}},(n-1)_l,n_{\bar{l}})
 = } & &
 \nonumber \\
 & &
 \sum\limits_{i=1}^{n-3} \overline{U}^{(0)}(1,...,i) V^\mu_{q,\bar{q},\gamma/Z} V^{(0)}(i+1,...,n-2)
 J^{\mathrm{EW}}_\mu(n-1,n).
\eq
Here, $V^\mu_{q,\bar{q},\gamma/Z}$ is the electroweak quark-photon/$Z$-boson vertex and
$J^{\mathrm{EW}}_\mu(n-1,n)$ the electroweak current, including the photon and the $Z$-boson propagator.

\subsection{One-loop recurrence relations}
\label{sect:loop_recursion}

We now turn to the computation of the integrand of the bare one-loop amplitude.
For a given loop momentum $k$ and external momenta $p_1$, ..., $p_n$ we consider the unintegrated one-loop currents
$J^{(1)}$ (unintegrated one-loop pure gluon current), 
$\overline{U}^{(1)}$ (unintegrated one-loop quark current)
and $V^{(1)}$ (unintegrated one-loop antiquark current).
Note that for a given $k$ and given $p_1$, ..., $p_n$ all momenta are fixed, in particular $k_j=k-q_j$, $q_j=p_1+...+p_j$.
\begin{figure}
\begin{center}
\includegraphics[bb=45 495 565 780,width=0.9\textwidth]{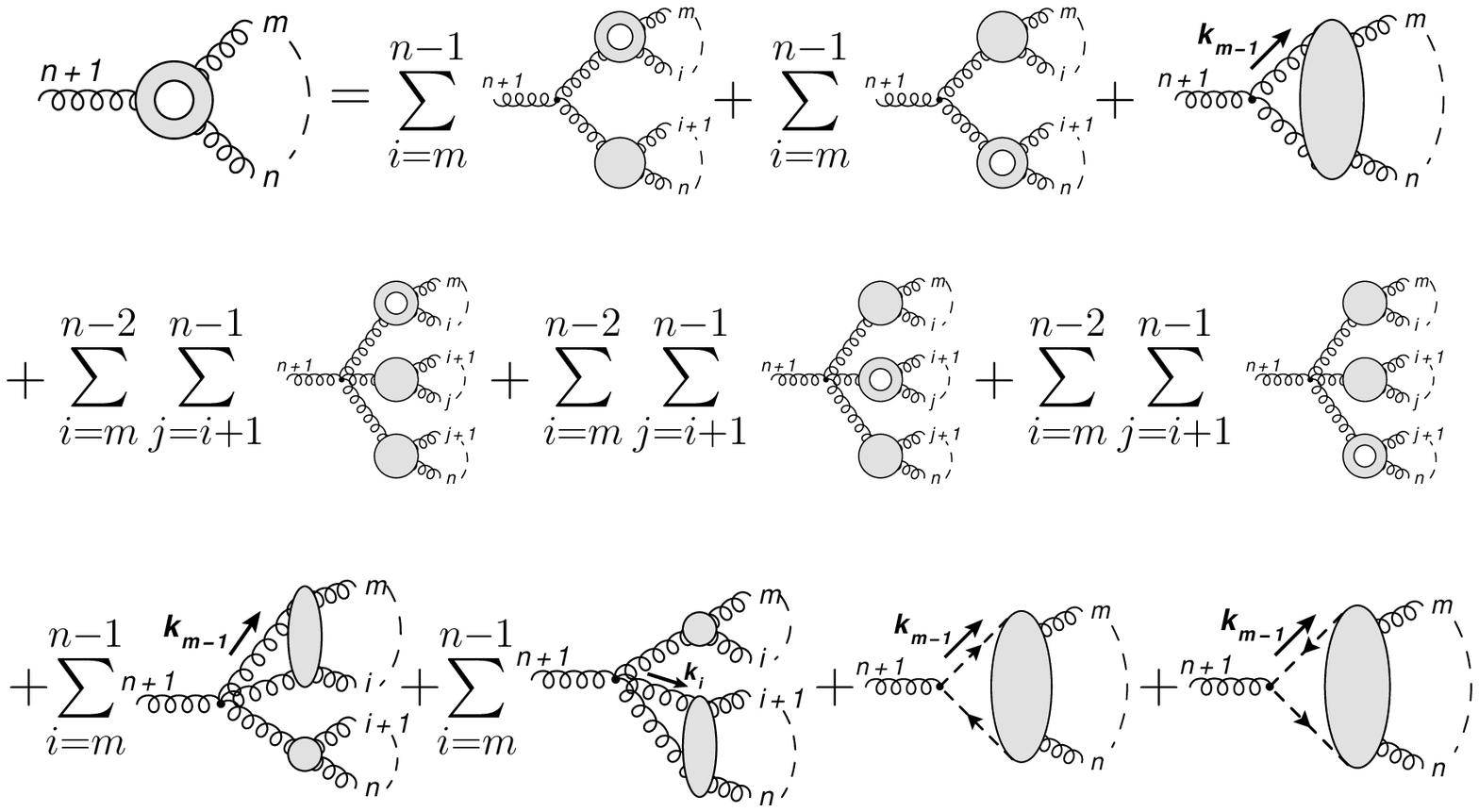}
\end{center}
\caption{\label{fig:oneloopgluon} The recurrence relation for the unintegrated one-loop gluon current.}
\end{figure}
\begin{figure}
\begin{center}
\includegraphics[bb=40 665 560 785,width=0.9\textwidth]{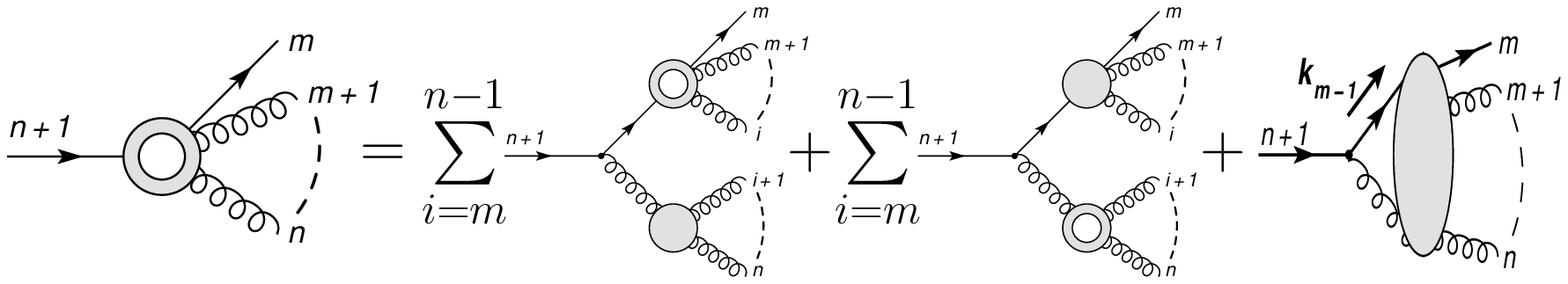}
\includegraphics[bb=40 685 560 785,width=0.9\textwidth]{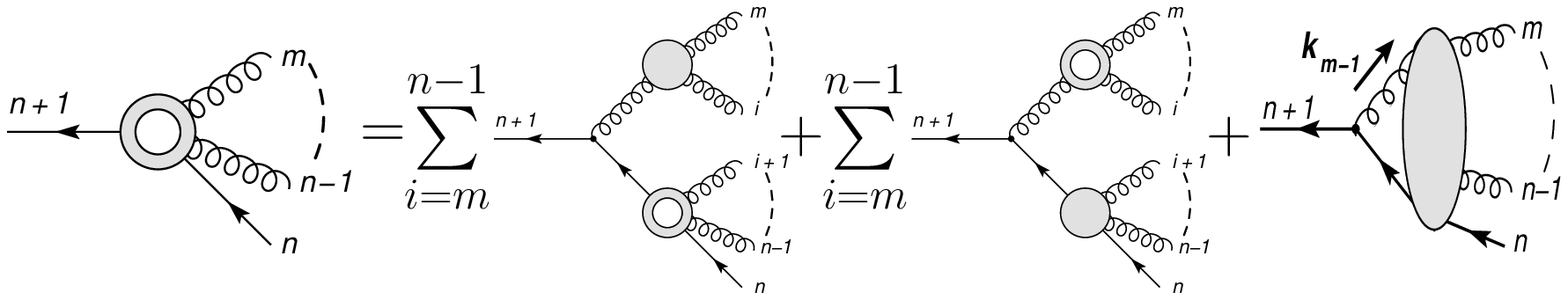}
\end{center}
\caption{\label{fig:oneloopquark} The recurrence relation for the unintegrated one-loop quark current and the 
unintegrated one-loop antiquark current.}
\end{figure}
The recursion relations for the unintegrated one-loop currents \cite{Draggiotis:2006er,vanHameren:2009vq} are shown in fig.~\ref{fig:oneloopgluon} 
for the gluon current
and in fig.~\ref{fig:oneloopquark} for the quark and the antiquark current.
In the recurrence relations for the unintegrated one-loop currents there are two types of vertices.
For the first type the off-shell leg couples to a tree-like vertex, 
in which case the recursion relates the unintegrated one-loop off-shell
current to an unintegrated one-loop off-shell current with fewer legs.
This recursion terminates with unintegrated one-loop one-currents, which are zero.
This corresponds to the fact that self-energy corrections on external lines are not included in the computation
of the integrand of the bare one-loop amplitude.

For the second type the off-shell leg couples directly through a vertex to the loop.
We call this contribution the ``direct contribution''.
In this case two edges of the vertex are connected to loop propagators. 
We can cut open one of these two edges by replacing the tensor structure of the corresponding propagator by a sum
over (pseudo-) polarisations.
Technically, this is done as follows: If the edge we would like to cut open corresponds to a gluon, we 
\begin{figure}
\begin{center}
\includegraphics[bb=45 695 565 780,width=0.9\textwidth]{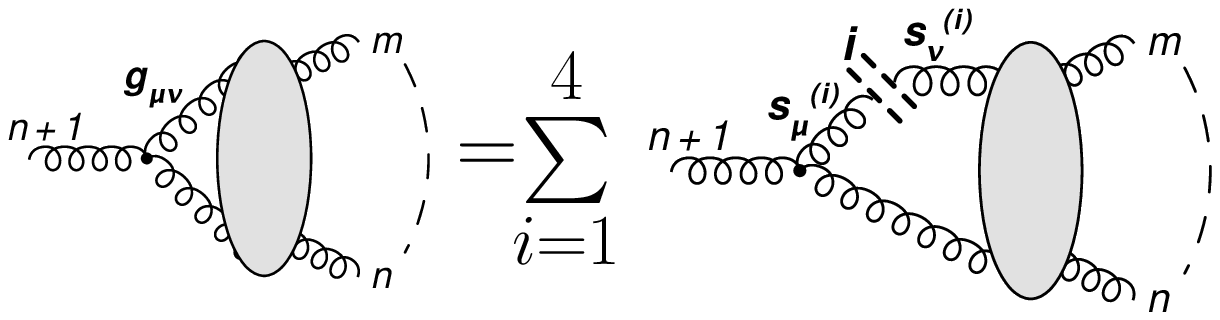}
\end{center}
\caption{\label{fig13} By cutting open the loop at a gluon line one replaces the tensor structure of the indicated
gluon loop propagator by a sum over four (pseudo-) polarisations.}
\end{figure}
replace $g_{\mu\nu}$ in the numerator of the gluon propagator in Feynman gauge by
\bq
     g_{\mu\nu} = \sum\limits_{i=1}^4 s^{(i)}_\mu s^{(i)}_\nu,
\eq
with four standard (pseudo-) polarisations
\bq
     s^{(1)}_\mu = \left( 1, 0, 0, 0 \right),
     & & 
     s^{(2)}_\mu = \left( 0, -i, 0, 0 \right),
 \nonumber \\
     s^{(3)}_\mu = \left( 0, 0, -i, 0 \right),
 & &
     s^{(4)}_\mu = \left( 0, 0, 0, -i \right).
\eq
This is shown in fig.~\ref{fig13}.
If, on the other hand, the edge corresponds to a (massless) quark line, we replace $k\!\!\!/$ in the numerator of 
the quark propagator by
\bq
\label{cut_quark1}
 k\!\!\!/ = k\!\!\!/^\flat + \frac{k^2}{2kq} q\!\!\!/,
 \;\;\;
 k^\flat = k - \frac{k^2}{2kq} q,
\eq
where $q$ is a light-like reference momentum and $k^\flat$ is by construction light-like.
\begin{figure}
\begin{center}
\includegraphics[bb=45 700 565 790,width=0.9\textwidth]{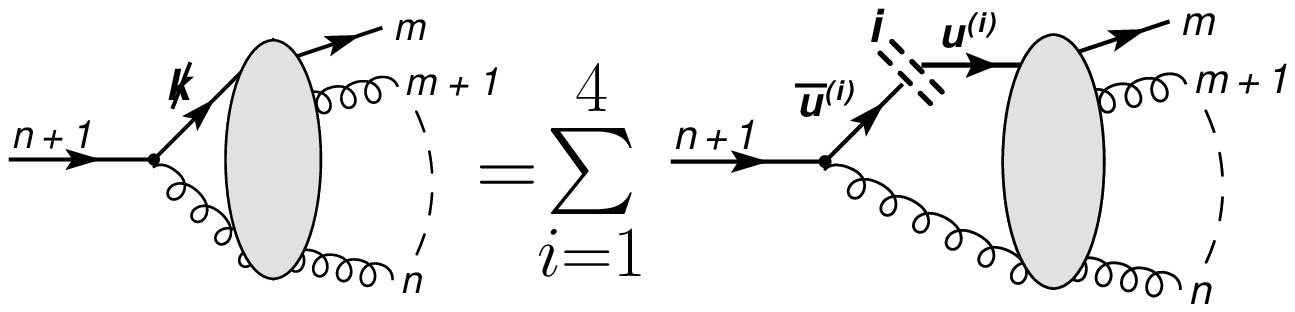}
\end{center}
\caption{\label{fig16} By cutting open the loop at a quark line one replaces the tensor structure of the indicated
quark loop propagator by a sum over four (pseudo-) polarisations.}
\end{figure}
We then replace $k\!\!\!/^\flat$ and $q\!\!\!/$ by a polarisation sum
\bq
\label{cut_quark2}
 k\!\!\!/^\flat & = & \sum\limits_{\lambda=\pm} u(k^\flat,\lambda) \bar{u}(k^\flat,\lambda),
 \nonumber \\
 q\!\!\!/ & = & \sum\limits_{\lambda=\pm} u(q,\lambda) \bar{u}(q,\lambda).
\eq
This is shown in fig.~\ref{fig16}.
The construction above for massless quarks can be generalised to massive quarks.

In the case of a one-loop gluon current we also have to cut open the ghost loop.
Since the ghosts are scalar particles, this is rather simple and does not involve a non-trivial polarisation sum.

As a sideremark we note that within the context of calculations based on Feynman diagrams the technique of cutting open the loop has 
recently been discussed in ref.~\cite{Cascioli:2011va}.

In all cases after cutting open the loop we obtain an object which we call a chain.
This object can again be calculated recursively. The recursion relation corresponds to a tree-like
calculation with $(n+1)$ external legs. One of the external legs corresponds to the cut loop propagator.
This leg is not on-shell. However this does not affect the recurrence relations.
\begin{figure}
\begin{center}
\includegraphics[bb=40 600 560 730,width=0.9\textwidth]{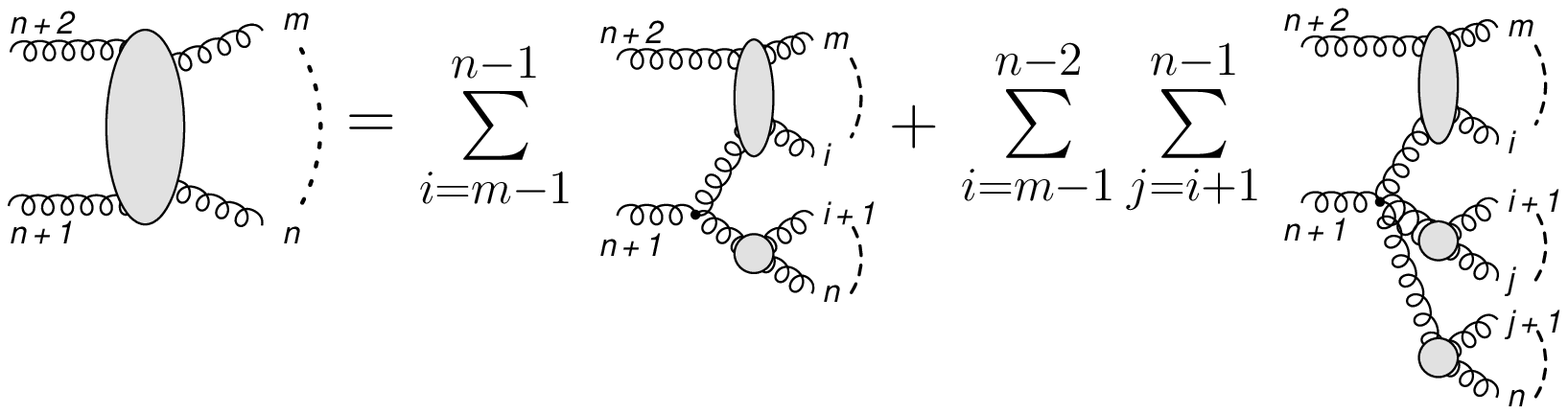}
\end{center}
\caption{\label{fig14} The recurrence relation for the the pure gluon chain. The chain results from cutting open the loop.}
\end{figure}
\begin{figure}
\begin{center}
\includegraphics[bb=40 655 560 755,width=0.9\textwidth]{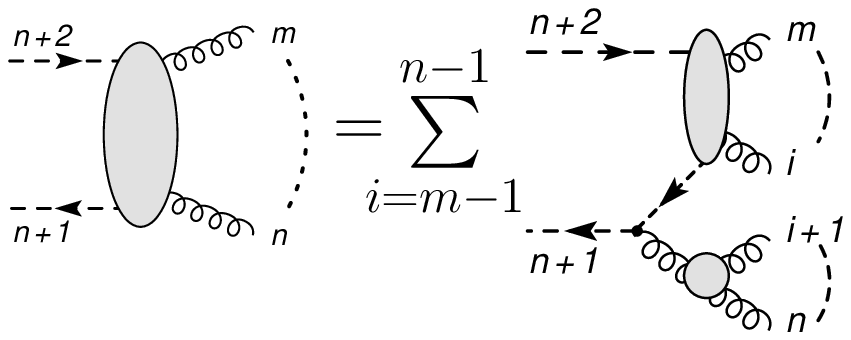}
\end{center}
\caption{\label{fig15} The recurrence relation for the the clockwise ghost-antighost chain. The chain results from cutting open the ghost loop.}
\end{figure}
We show two examples for the recurrence relations of chains.
In fig.~\ref{fig14} the recurrence relation for the pure gluon chain is given.
As a second example we show in fig.~\ref{fig15} the recurrence relation for the clockwise
ghost-antighost chain.
In addition there are further chains:
From cutting open all direct contributions to the unintegrated one-loop gluon current, we also obtain the anti-clockwise ghost-antighost chain.
Cutting the direct contribution to the unintegrated one-loop quark current gives a
quark-gluon chain, cutting the corresponding contribution to the unintegrated one-loop antiquark current
gives a gluon-antiquark chain.
These chains have similar recurrence relations, which we do not show explicitly.

For completeness we mention that for the process $e^+ e^- \rightarrow q, g, ..., g, \bar{q}$ we also need the
unintegrated one-loop electroweak quark-gluon-antiquark current.
\begin{figure}
\begin{center}
\includegraphics[bb=40 685 560 785,width=0.9\textwidth]{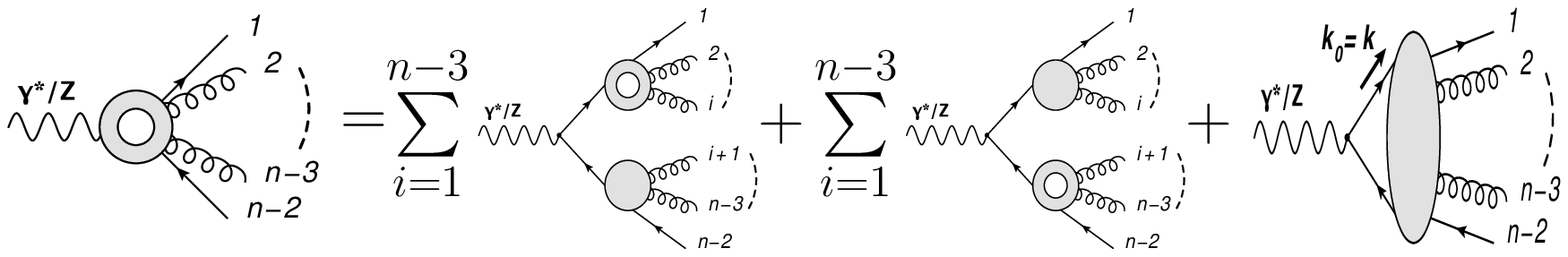}
\end{center}
\caption{\label{fig8} The recurrence relation for the unintegrated one-loop electroweak quark-gluon-antiquark current.}
\end{figure}
This current can be calculated with the same methods as discussed above.
The recurrence relation for this current is shown in fig.~\ref{fig8}.
In the direct contribution we cut open the quark propagator as outlined in eq.~(\ref{cut_quark1})
and eq.~(\ref{cut_quark2}).
We obtain an antiquark-quark chain, which again can be computed with a tree-like recurrence relation.

\subsection{UV recurrence relations}
\label{sect:uv_recursion}

Let us now turn to the computation of the ultraviolet subtraction term.
We recall that the basic building blocks of the ultraviolet subtraction term are propagator and
vertex subtraction terms.
We can think of the ultraviolet subtraction term
as a sum over diagrams, where each diagram has a tree-structure with exactly one propagator or vertex replaced by a
basic ultraviolet subtraction term.
The complete ultraviolet subtraction term can again be calculated recursively.
\begin{figure}
\begin{center}
\includegraphics[bb=50 495 560 780,width=0.9\textwidth]{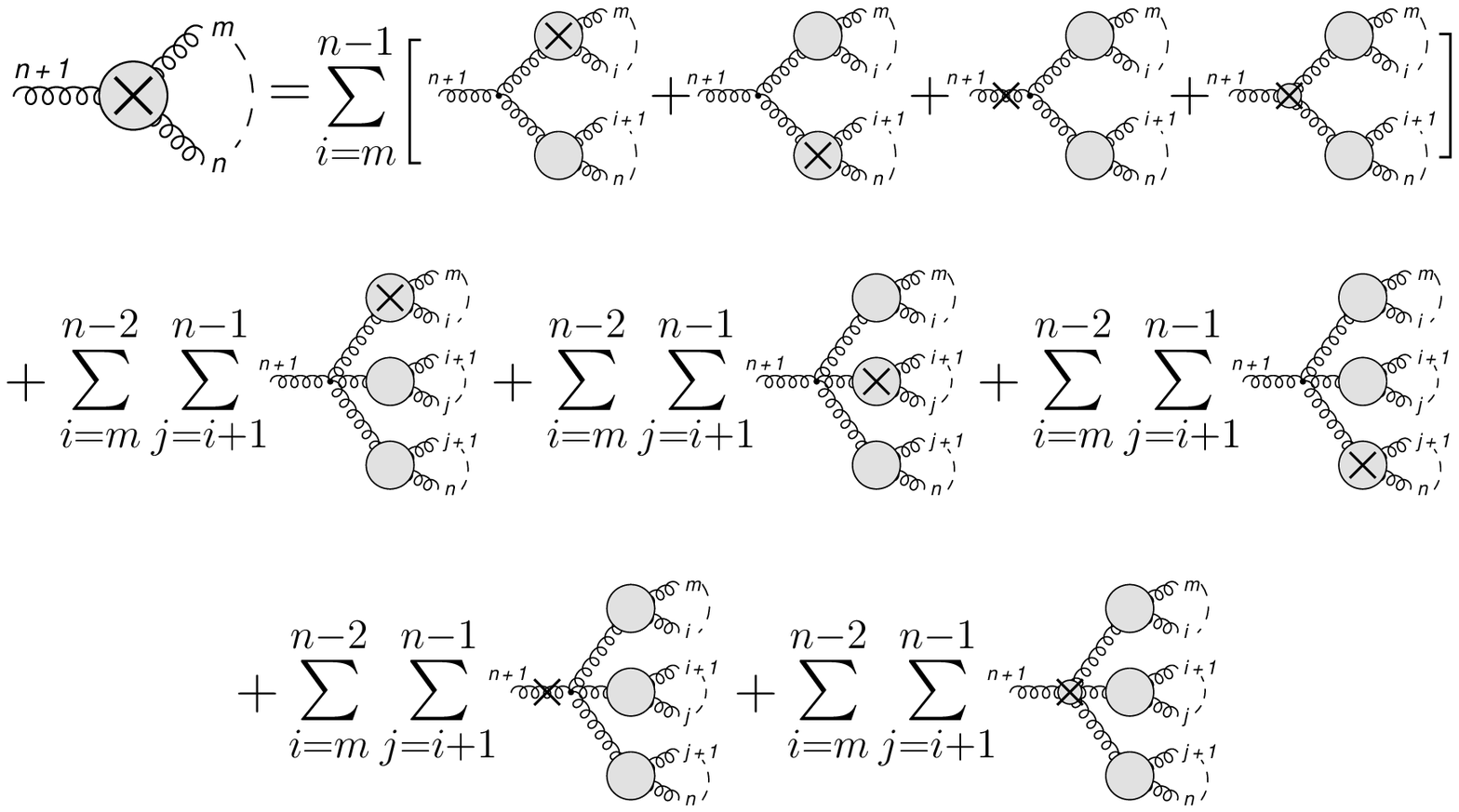}
\end{center}
\caption{\label{fig10} The recurrence relation for the ultraviolet subtraction term to the pure gluon current.}
\end{figure}
The recursion relation is shown in fig.~\ref{fig10} for the gluon current 
and in fig.~\ref{fig9} for the quark current.
\begin{figure}
\begin{center}
\includegraphics[bb=50 525 560 780,width=0.9\textwidth]{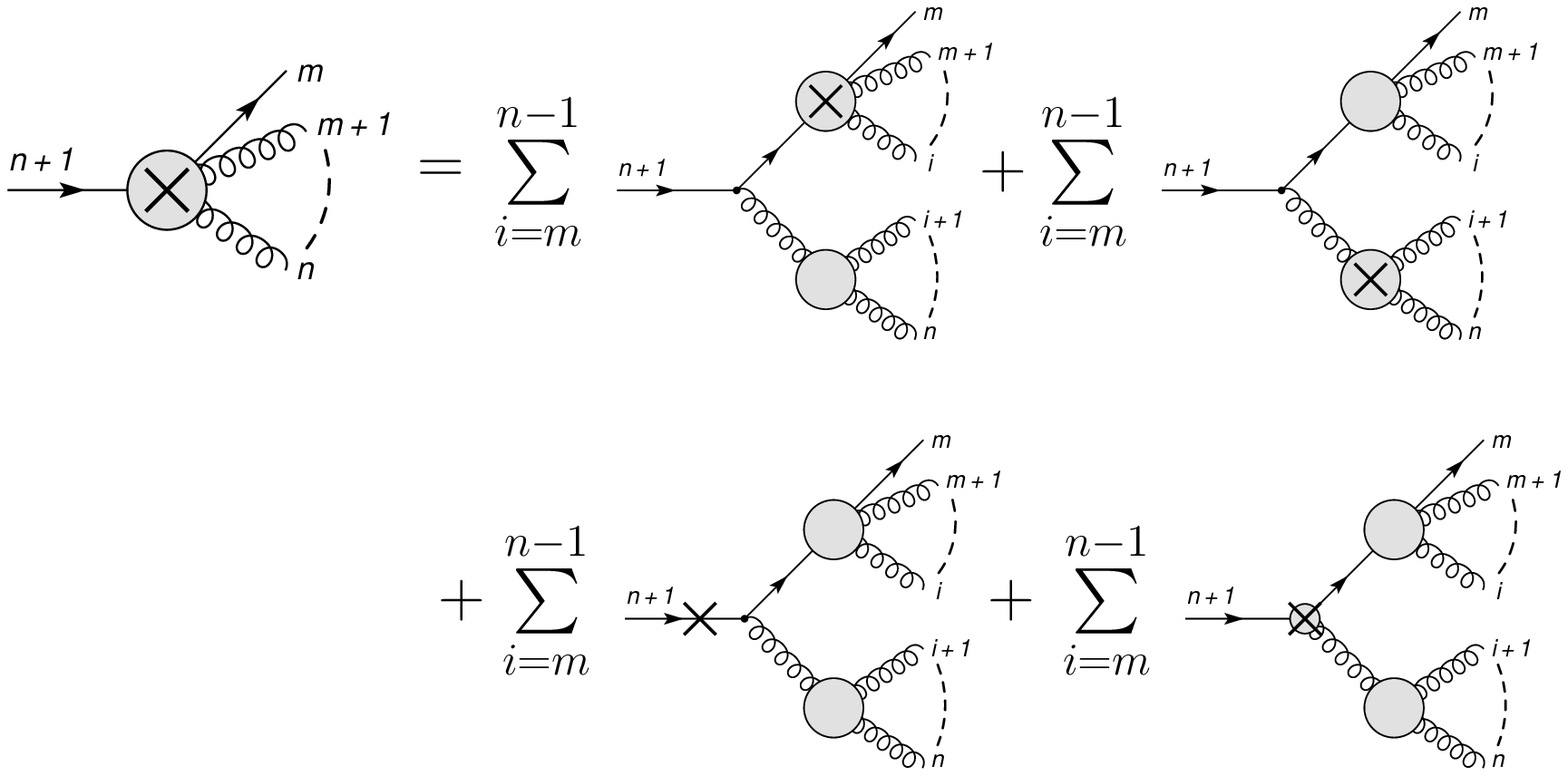}
\end{center}
\caption{\label{fig9} The recurrence relation for the ultraviolet subtraction term to the quark current.}
\end{figure}
The recurrence relation for the antiquark current has a similar structure as the recurrence relation for the quark current
and is not shown explicitly. 
The structure of the recursion relations is as follows:
Either the off-shell leg is attached through a Born propagator to a Born vertex. In this case exactly one of the sub-currents
carries a basic ultraviolet subtraction term, while all other sub-currents are Born currents.
Or the off-shell leg is connected through a basic ultraviolet subtraction term to the sub-currents.
The basic ultraviolet subtractions terms can be either a propagator subtraction term or a vertex subtraction term.
In this case all sub-currents are Born currents.

For the process $e^+ e^- \rightarrow q, g, ..., g, \bar{q}$ we also need the
off-shell current including the ultraviolet subtraction term for the electroweak quark-gluon-antiquark vertex.
\begin{figure}
\begin{center}
\includegraphics[bb=40 685 560 785,width=0.9\textwidth]{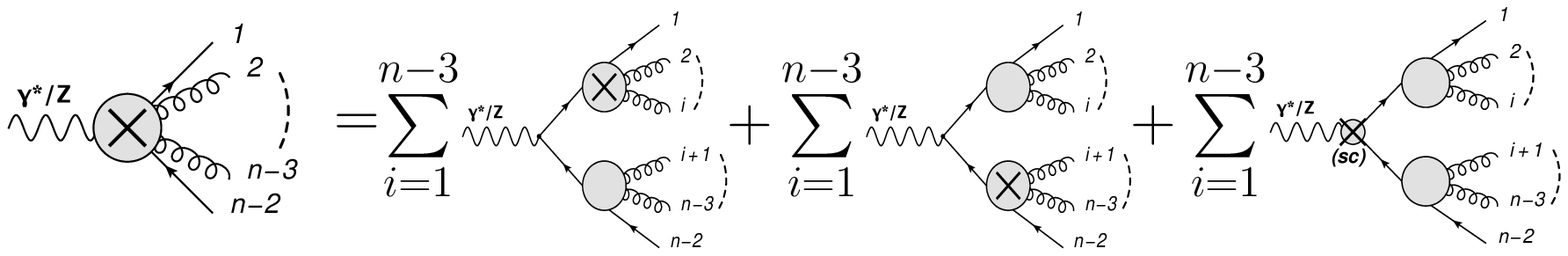}
\end{center}
\caption{\label{fig11} The recurrence relation for the ultraviolet subtraction term to the electroweak
quark-gluon-antiquark current.}
\end{figure}
The corresponding recursion relation is shown in fig.~\ref{fig11}.
The basic ultraviolet subtraction term for the electroweak quark-gluon-antiquark vertex
can be obtained from the sub-leading $N_c$ quark-gluon vertex subtraction term by adjusting the coupling factor
appropriately.


\section{Conclusions}
\label{sect:conclusions}

In this paper we have given a detailed account on the techniques used to improve the efficiency of the
Monte Carlo integration in a numerical approach for the computation of one-loop QCD amplitudes.
The techniques fall into two categories:
Techniques in the first category reduce the statistical error of the Monte Carlo integration.
This is done by dividing the integration into sub-channels and optimising the integration in each sub-channel.
Of particular importance is the improvement of the ultraviolet subtraction beyond the formally required
$|k|^{-5}$ fall-off.
Within the second category we discussed CPU-efficient methods for the computation of the integrands.
This is done with the help of recurrence relations.

All techniques discussed in this paper are process-independent or have an obvious generalisation towards
more complex processes.

\bibliography{/home/stefanw/notes/biblio}
\bibliographystyle{/home/stefanw/latex-style/h-physrev5}

\end{document}